\documentclass[structabstract]{aa}  
\usepackage{graphicx}
\usepackage{txfonts}
\usepackage[authoryear]{natbib}
\bibpunct[; ]{(}{)}{,}{a}{}{;}
\usepackage{longtable,lscape}
\begin{document}
   \title{NGC~7419 as a template for red supergiant clusters\thanks{Partially based on observations collected at the Nordic Optical Telescope and the William Herschel Telescope (La Palma) and at the 1.93-m telescope at Observatoire de Haute Provence (CNRS), France.}}

   \author{A. Marco\inst{1}
           \and
           I. Negueruela\inst{1}
           }


   \institute{Departamento de F\'{i}sica, Ingenier\'{i}a de Sistemas y Teor\'{i}a de la Se\~{n}al. Escuela Polit\'ecnica Superior. University of Alicante. Apdo.99 E-03080. Alicante. (Spain)\\
              \email{amparo.marco@ua.es}}

   \date{Received; accepted}
   \titlerunning{NGC 7419 and red supergiants}

 
  \abstract
   {The open cluster NGC~7419 is known to contain five red supergiants and a very high number of Be stars. However, there are conflicting reports about its age and distance that prevent a useful comparison with other clusters.}
   {We intend to obtain more accurate parameters for NGC~7419, using techniques different from those of previous authors, so that it may be used as a calibrator for more obscured clusters.}
   {We obtained Str\"{o}mgren photometry of the open cluster NGC~7419, as well as classification spectroscopy of $\sim20$ stars in the area. We then applied standard analysis and classification techniques.}
   {We find a distance of $4\pm0.4$ kpc and an age of $14\pm2$ Myr for NGC~7419. The main-sequence turn-off is found at spectral type B1, in excellent agreement. We identify 179 B-type members, implying that there are more than 1200 $M_{\sun}$ in B stars at present. Extrapolating this to lower masses indicates an initial cluster mass of between 7000 and 10000 $M_{\sun}$, depending on the shape of the Initial Mass Function. We find a very high fraction ($\approx40$\%) of Be stars around the turn-off, but very few Be stars at lower masses. We also report for the first time a strong variability in the emission characteristics of Be stars. We verified that the parameters of the red supergiant members can be used to obtain accurate cluster parameters.} 
   {NGC~7419 is sufficiently massive to serve as a testbed for theoretical predictions and as a template to compare more obscured clusters. The distribution of stars above the main-sequence turn-off is difficult to accommodate with current evolutionary tracks. Though the presence of five red supergiants is marginally consistent with theoretical expectations, the high number of Be stars and very low number of luminous evolved B stars hint at some unknown physical factor that is not considered in current synthesis models.}

   \keywords{open cluster and associations: individual: NGC 7419 - stars:
             Hertzsprung-Russell (HR) and C-M diagrams, early-type, evolution, 
             emission-line, Be, supergiant - techniques:photometric and 
             spectroscopic
               }

   \maketitle
%

\section{Introduction}

Open clusters are excellent laboratories for the study of stellar evolution, where we can find rare evolutionary phases in their natural environment. In addition, most massive stars are found in open clusters, because their short lifetimes prevent them from drifting into the field. Unfortunately, the most massive (and interesting) clusters are usually affected by very heavy foreground extinction, makes it difficult and very expensive in telescope time to study them. Very detailed study of accessible open clusters can provide us with valuable information to interpret the much poorer datasets available for very heavily reddened clusters. 

NGC~7419 is an open cluster located in Cepheus, near the Galactic plane ($l$=109$\degr$, $b$=1$\fdg14$). Its spatial extent is very well defined, with a strong concentration towards the centre and little foreground contamination \citep{joshi2008}. \citet{beauchamp1994} estimated its distance, age, and mass function, based on $UBV$ CCD photometry and a comparison with the isochrones of \citet{maeder1990}. They obtained a distance of 2.3~kpc, an age of $14\pm2$~Myr, and a mass function that satisfies the Salpeter form. Their photometry was not calibrated with standard stars, but with the photometric sequence of the nearby cluster NGC~7510. \citet{turner1983} had noted that the same procedure had led to large systematic differences in $(U-B)$ in the study of the nearby open cluster Mrk~50. 
 
More recently, \citet{subramanian2006} also presented $UBV$ photometry of the cluster, finding an age of $25\pm5$~Myr and a distance of 2.9~kpc. \citet{joshi2008} estimated, from $UBVRI$ photometry,  an age of $23\pm3$~Myr and a distance of 3.3~kpc.  
Unfortunately, there are very large systematic differences between the photometric values reported in these three works. \citet{joshi2008} and \citet{subramanian2006} present substantial differences in $(U-B)$ and $(B-V)$ with respect to \citet{beauchamp1994}, which may have some dependence on the observed colours. There are, moreover, important differences between \citet{joshi2008} and \citet{subramanian2006}, which do not seem to depend on the observed colours, but reach $\sim0.2$~mag in $(U-B)$ for many stars. The reasons for these differences are unclear. They may be related to the high reddening towards the cluster, perhaps combined with the choice of calibration. Indeed, of the three works, \citet{joshi2008} is the only one to calibrate the photometry with Landolt standards.

Based on photometric searching techniques, \citet{pigulski2000} determined the fraction of Be stars to total number of B-type stars to be $36\pm7\%$ among cluster members brighter than $R_{C}=16.1$~mag. Such a high fraction of Be stars places NGC~7419 in the small group of clusters very rich in Be stars, such as NGC~663 in the Milky Way, NGC~330 in the Small Magellanic Cloud (SMC), and NGC~1818A in the Large Magellanic Cloud (LMC). This high fraction of Be stars has been confirmed by \citet{subramanian2006} using slitless spectroscopy and by \citet{joshi2008} using H$\alpha$ photometry. \citet{malchenko2011} obtained intermediate-resolution spectroscopy around the H$\alpha$ line for most of the candidate Be stars, confirming their nature.

A second peculiarity of NGC~7419 is the presence of five red supergiants (RSGs) as confirmed radial velocity members of the cluster \citep{beauchamp1994}. Until recently, this was the highest number of RSGs in a Galactic cluster \footnote{There are five RSGs associated with NGC~884, but four of them are located in the extended halo or association surrounding the cluster \citep{mermilliod2008}.}. \citet{beauchamp1994} and \citet{caron2003} noticed that this high number of RSGs was not accompanied by any blue supergiant. This proportion is at odds with what is generally found in Galactic open clusters, where blue supergiants outnumber RSGs by a factor $\sim3$ \citep{eggenberger2002}.

In recent years, several clusters of RSGs have been discovered towards the Scutum Arm tangent \citep{figer2006, davies2007, clark2009a}. These clusters are so heavily obscured that their population of intrinsically blue stars have not been found yet, and only the RSGs have been observed as bright sources in the near-infrared. Population synthesis models indicate that, given the rarity of RSGs, the clusters must contain very large stellar populations. Simulations by \citet{clark2009b} predict a stellar mass of about $10^{4}$ for each 1--3 RSGs at 10 Myr and 3--5 RSGs at 20 Myr. NGC~7419 offers the opportunity of studying a cluster with a moderately high number of RSGs whose population of B-type stars can be characterised in detail.       

In this paper, we determine the fundamental parameters of NGC~7419 using techniques different and more accurate than those of previous authors: Str\"{o}mgren photometry and intermediate-resolution spectroscopy in the classification region. In view of the very discrepant photometric measurements reported in previous studies, we have decided to try intermediate-band Str\"omgren photometry which, if correctly calibrated, may allow easy individual dereddening of stellar colours and give accurate photometric spectral types. Intermediate-resolution spectroscopy allows spectral classification that serves to test the accuracy of dereddening procedures and photometric results.

The paper is organised as follows. In Section~\ref{obs}, we describe the photometric and spectroscopic runs in which data were gathered for this work. In Sections~\ref{class} and~\ref{RSGs}, we derive spectral types for the brightest blue members of the cluster and the red supergiant members, while in Section~\ref{bes} we present evidence for variability in Be stars. In Section~\ref{HR}, we use our Str\"omgren photometry to estimate the cluster parameters, and then calculate the cluster mass in Section~\ref{mass}. Finally, in Section~\ref{dis}, we compare our results to those of previous studies and interpret them in view of theoretical expectations and comparison to other clusters, highlighting the role of NGC~7419 as a template for more obscured clusters. 

\section{Observations and data}
\label{obs}

In this work, we present Str\"omgren photometry and multi-epoch spectroscopy of NGC~7419 for the first time. 
Seven different campaigns of spectroscopic observations have been carried out over the past ten years.  
We began in 2001 by taking slit spectra of stars catalogued as Be by \citet{pigulski2000} with the 1.93-m telescope at the Haute Provence Observatory (France) and the Carelec spectrograph. In 2004, we had an observation run to take photometry and spectroscopy with the 2.6-m Nordic Optical Telescope (NOT, La Palma, Spain). The telescope was equipped with the imager and spectrograph ALFOSC. On this occasion, the weather conditions were not photometric and we could only take slit spectra of the brightest stars and slitless images to detect emission-line stars. In 2005, we had another run at the same telescope. On that occasion, weather conditions were excellent and we could take photometric observations. In addition, we took more slit spectra of bright members. 
Spectra of the red supergiant members of the cluster were taken in 2007 and 2009 with the William Herschel Telescope and ISIS. 
In 2008 and 2011, we obtained slitless images again to detect emission-line stars with the NOT, and in 2010 we took a few more slit spectra. 
With this large collection of data, we are able to confirm the extremely high fraction of Be stars, find evidence of variability in many of them, and determine spectral types for the brightest members.

\subsection{Photometry}

We obtained Str\"{o}mgren photometry of the open cluster NGC~7419 using ALFOSC on the Nordic Optical Telescope at the Roque de los Muchachos Observatory (La Palma, Spain) on the night of October 3, 2005. ALFOSC allows observations in different modes. In imaging mode the camera covers a field of $6\farcm5 \times 6\farcm5$ and has a pixel scale of $0\farcs19/$ pixel. For each frame, we obtained two series of different exposure times in each filter to achieve accurate photometry for a broad magnitude range. The central position for the cluster and the exposure times are presented in Table~\ref{tab1}. 

\begin{table}
\caption{Log of the photometric observations taken at the NOT in October 2005. \label{tab1}}
\centering
\begin{tabular}{lll}
\hline\hline
\noalign{\smallskip}
Name&RA(J2000)&Dec(J2000)\\
\hline
\noalign{\smallskip}
 NGC~7419& 22h 54m 18.7s & $+60\degr 48\arcmin 49\arcsec$ \\
NGC~6910& 20h 23m 10.6s &  $+40\degr 46\arcmin 22.4\arcsec$ \\
NGC~1502& 04h 07m 40.4s &  $+62\degr 20\arcmin 59.1\arcsec$ \\
NGC~~869&02h 19m 4.4s &  $+57\degr 08\arcmin 7.8\arcsec$ \\
NGC~~884&02h 22m 0.6s &  $+57\degr 08\arcmin 42.1\arcsec$ \\
\hline
\end{tabular}
\centering
\begin{tabular}{l c c}
\noalign{\smallskip}
\multicolumn{3}{c}{Exposure times(s)}\\
Filter & Long times & Short times \\
\hline
\noalign{\smallskip}
\multicolumn{3}{c}{NGC~7419}\\
\hline
\noalign{\smallskip}
$u$ & 1800 & 400  \\
$v$ & 1000 &200 \\
$b$ & 600 &120  \\
$y$ & 120 & 30 \\
\hline\hline
\noalign{\smallskip}
\multicolumn{3}{c}{NGC~6910}\\
\hline
\noalign{\smallskip}
$u$ & 30 & 10  \\
$v$ & 12 &5 \\
$b$ & 10 &3  \\
$y$ & 5& 2 \\
\hline\hline
\noalign{\smallskip}
\multicolumn{3}{c}{NGC~1502}\\
\hline
\noalign{\smallskip}
$u$ & $-$& 10  \\
$v$ & $-$ &5 \\
$b$ & $-$ &3  \\
$y$ & $-$ & 2 \\
\hline\hline
\noalign{\smallskip}
\multicolumn{3}{c}{NGC~869 \& NGC~884}\\
\hline
\noalign{\smallskip}
$u$ & $-$&5   \\
$v$ & $-$ &3 \\
$b$ & $-$ &1  \\
$y$ & $-$ &1 \\
\hline\hline
\end{tabular}
\tablefoot{Cluster coordinates are taken from the WEBDA database ({\tt http://www.univie.ac.at/webda}).}
\end{table}

Standard stars were observed the same night in the clusters NGC~869, NGC~884, NGC~6910, and NGC~1502 using exposure times suitable to obtain good photometric values for all stars selected as standards in these clusters. The central positions for the clusters and the exposure times are presented in Table~\ref{tab1}.

The choice of standard stars for the transformation
is a critical issue in $uvby$ photometry \citep{delgado1989}. Transformations
made only with unreddened stars introduce large systematic
errors when applied to reddened stars, even if the
colour range of the standards brackets that of the programme
stars \citep{manfroid1987,crawford1994}.

We have been carrying out CCD Str\"{o}mgren photometry of open clusters for years and have always been limited by the scarcity of suitable secondary standard stars. To resolve this problem, we built a list of non-variable and 
non-peculiar (at least, not reported as such in the literature) candidate stars in different clusters observed with the same Kitt Peak telescopes and instrumentation used to define the $uvby$ \citep{crawford1970a} and H$\beta$ \citep{crawford1966} standard systems, so that there is no doubt that the photometric values are in the standard systems.

A preliminary list of standard stars was presented in \citet{marco2001}, but dedicated searches for variable stars in the selected open clusters have resulted in the identification of several new variables \citep{huang2006, saesen2010a, saesen2010b}.
After cleaning the original list, we provide in Table~\ref{tab2} a selection of standards that are almost certainly not variable. The list of adopted standard stars and their photometric
data to be used in the transformations of CCD Str\"omgren photometry of open clusters are given in Table~\ref{tab2}. In this run we used only standard stars in NGC~6910, NGC~1502, NGC~884, and NGC~869.

The bias and flat-field corrections of all frames was performed with IRAF\footnote{IRAF is distributed by the National Optical Astronomy Observatories, which are operated by the Association of Universities for Research in Astronomy, Inc., under cooperative agreement with the National Science Foundation} routines. Photometry was obtained by
point-spread function (PSF) fitting using the DAOPHOT package
\citep{stetson1987} provided by IRAF. The apertures used are of the order of 
the full width at half maximum (FWHM). In this case, we used a value of three pixels for each image in all filters. To construct the PSF empirically,
we automatically selected bright stars (typically 25 stars). After this,
we reviewed the candidates and discarded those not fulfilling the
requirements for a good PSF star. Once we had a list of PSF
stars ($\approx 20$), we determined an initial PSF by fitting the best
function of the five options offered by the PSF routine inside
DAOPHOT. We allowed the PSF to be variable (in order 2) across the
frame to take into account the systematic pattern of PSF variability
with the position on the chip. We needed to perform an aperture correction of 15 pixels for each frame in all filters.

The atmospheric extinction corrections were performed using the RANBO2 program, which implements the method described by \citet{manfroid1993}. Finally, we obtained the instrumental magnitudes for all stars.

\begin{table*}
\centering
\caption{Standard stars with their catalogued values and spectral types taken from the literature.\label{tab2}}
\begin{tabular}{ccccccc}
\hline
\noalign{\smallskip}
Star&$V$&$b-y$&$m_{1}$&$c_{1}$&$H\beta$&spectral type\\
\hline
\multicolumn{7}{c}{NGC~869}\\
\hline
\noalign{\smallskip}
782 &  9.450   &   0.268 &  -0.053 &   0.150 & 2.618 &- \\
800 &  12.250  &   0.316 &  -0.031 &   0.404 & 2.701 &-\\
837 &  14.080  &   0.393 &   0.000 &   0.918 & 2.801 &- \\
867 &  10.510  &   0.393 &   0.161 &   0.375 &2.613  &- \\
935 &  14.020  &   0.362 &  -0.004 &   0.854 &2.802  &- \\
950 &  11.290  &   0.318 &  -0.048 &   0.214 &2.642  & B2V\\
978 &  10.590  &   0.305 &  -0.039 &   0.177 &2.643  & B2V-B1.5V\\
980 &  9.650   &   0.290 &  -0.054 &   0.167 & - &B1.5V \\
991 &  11.320  &   0.329 &  -0.062 &   0.275 & - &B2V \\
1015&  10.570  &   0.225 &   0.033 &   0.741 &2.772  & B8V\\
1078&  9.750   &   0.316 &  -0.065 &   0.167 &2.610  & B1V\\
1181&  12.650  &   0.372 &  -0.034 &   0.379 &2.718  & -\\
1187&  10.820  &   0.348 &  -0.063 &   0.212 &2.648  & B2IV\\
\hline
\noalign{\smallskip}
\multicolumn{7}{c}{NGC~884}\\
\hline
\noalign{\smallskip}
2005 & 13.560	& 0.35  & -0.013 &  0.713 & 2.750 & -  \\ 
2139 & 11.380	& 0.255 & -0.033 &  0.196 &2.649  &B2V	\\
2147 & 14.340	& 0.406 & -0.050 &  1.002 &2.863  &-	\\
2167 & 13.360	& 0.352 & -0.056 &  0.627 &2.752  &-	\\
2196 & 11.570	& 0.250 & -0.006 &  0.210 &2.670  &B1.5V\\
2235 & 9.360	& 0.316 & -0.088 &  0.150 & 2.611 &-	\\
2330 & 11.420	& 0.267 & -0.050 &  0.277 &2.640  &-	\\
2572 & 10.020	& 0.311 & -0.091 &  0.197 &2.620  &-	\\
\hline
\noalign{\smallskip}
\multicolumn{7}{c}{NGC~6910}\\
\hline
\noalign{\smallskip}
10 & 10.750 & 0.100&  0.150 & 1.040 & 2.909& -\\
20 & 10.900 & 0.770&  0.420 & 0.430 &2.555 &- \\
24 & 11.720 & 0.660& -0.140 & 0.220 &2.647 & B1V\\
21 & 11.730 & 0.590& -0.100 & 0.220 &2.652 & B1V\\
28 & 12.220 & 0.590& -0.110 & 0.330 &2.679 & -\\
172?& 13.900 & 0.710& -0.080 & 0.590 & & \\
\hline
\multicolumn{7}{c}{NGC~1502}\\
\hline
30 &  9.650 & 0.475 & -0.096 & 0.157 & 2.628 &B1~V\\
35 &  10.460& 0.422 & -0.052 & 0.265 & 2.661 &B3~V  \\
41 &  13.030& 0.552 &  0.003 & 1.232 & 2.874 &  \\
42 &  12.590& 0.510 & -0.078 & 0.960 & 2.770 &  \\
45 &  11.460& 0.475 & -0.065 & 0.526 & 2.723 &  \\
49 &  10.710& 0.436 & -0.058 & 0.316 & 2.656 &B3~V  \\
\hline
\noalign{\smallskip}
\multicolumn{7}{c}{NGC~1039}\\
\hline
\noalign{\smallskip}
267& 11.960& 0.303& 0.138 &0.481& 2.678&      \\ 
274& 9.740 & 0.086& 0.176 &0.973& 2.890&  A2.5V    \\
278 & 11.820& 0.144& 0.198 &0.900& 2.855&      \\
284& 10.760& 0.151& 0.194 &0.894& 2.848&      \\
294& 11.220& 0.176& 0.204 &0.796& 2.817&      \\
301& 10.030& 0.066& 0.152 &1.013& 2.916&      \\
303& 9.950 & 0.055& 0.163 &1.021& 2.908&       \\
\hline
\end{tabular}
\tablefoot{Photometric data are taken from \citet{crawford1970b} and \citet{johnson1955} for NGC~869 and NGC~884; \citet{perry1978} for NGC~2169; \citet{crawford1977} for 
NGC~6910; \citet{canterna1979} for NGC~1039, and \citet{crawford1994} for NGC~1502. Spectral types, when available, are taken from \citet{schild1965}, \citet{Slettebak1968} and \citet{johnson1955} for NGC~869 and NGC~884; \citet{perry1978} for NGC~2169; \citet{morgan1956} and \citet{hoag1965} for NGC~6910; \citet{abt1977} for  NGC~1039, and \citet{hoag1965} for NGC~1502.}                
\end{table*}
     
Using the standard stars, we transformed the instrumental magnitudes to the
standard system using the PHOTCAL package inside IRAF. We implemented the following transformation equations, after Crawford \& Barnes (1970a):

\begin{eqnarray}
V& = &15.565-0.021(b-y)+y_{i}\\
&&\pm0.011\pm0.020 \nonumber \\
(b-y)& = &1.022 + 0.870(b-y)_{i}\\
   &&  \pm0.014  \pm0.024 \nonumber \\
c_{1}& = &0.362 + 0.933{c_{1}}_{i} - 0.040(b-y)  \\
    && \pm0.025\pm0.029\pm0.043\, , \nonumber
\end{eqnarray}

where each coefficient is given with the error resulting from the transformation.

\begin{figure*}[ht]
\resizebox{12 cm}{!}{\includegraphics[angle=-90]{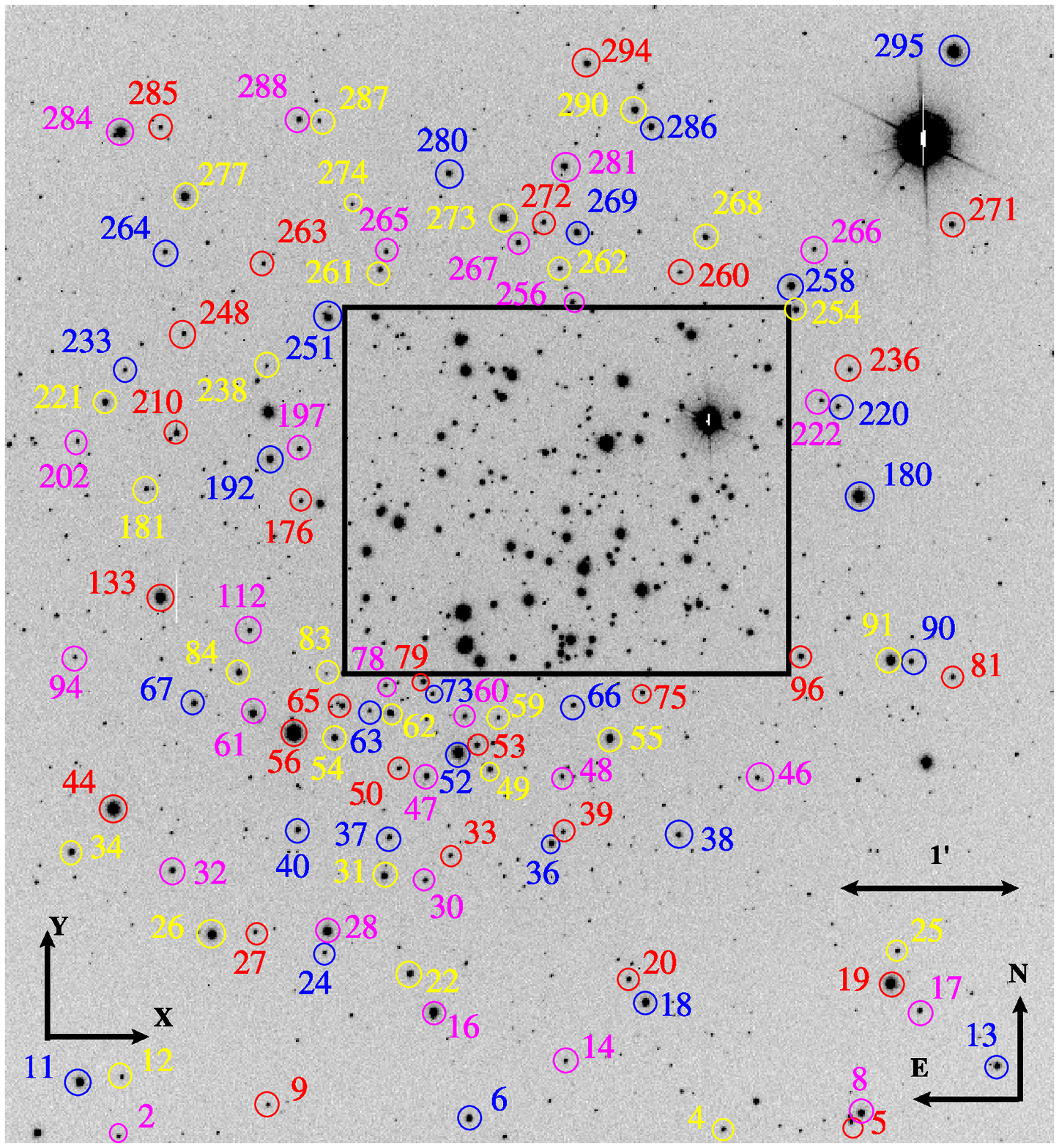}}
\centering
\caption{Finding chart for stars with photometry in the field of NGC~7419. The image is one of our $V$-band frames. Stars inside the rectangle (which approximately defines the cluster core) are marked in Fig.~\ref{Fig2}. Each star is identified by the nearest marker in the same colour as the circle around it. The origin of the coordinates is located at the bottom-left corner of each frame.} 
\label{Fig1}
\end{figure*}

\begin{figure*}[ht]
\resizebox{16 cm}{!}{\includegraphics[angle=-90]{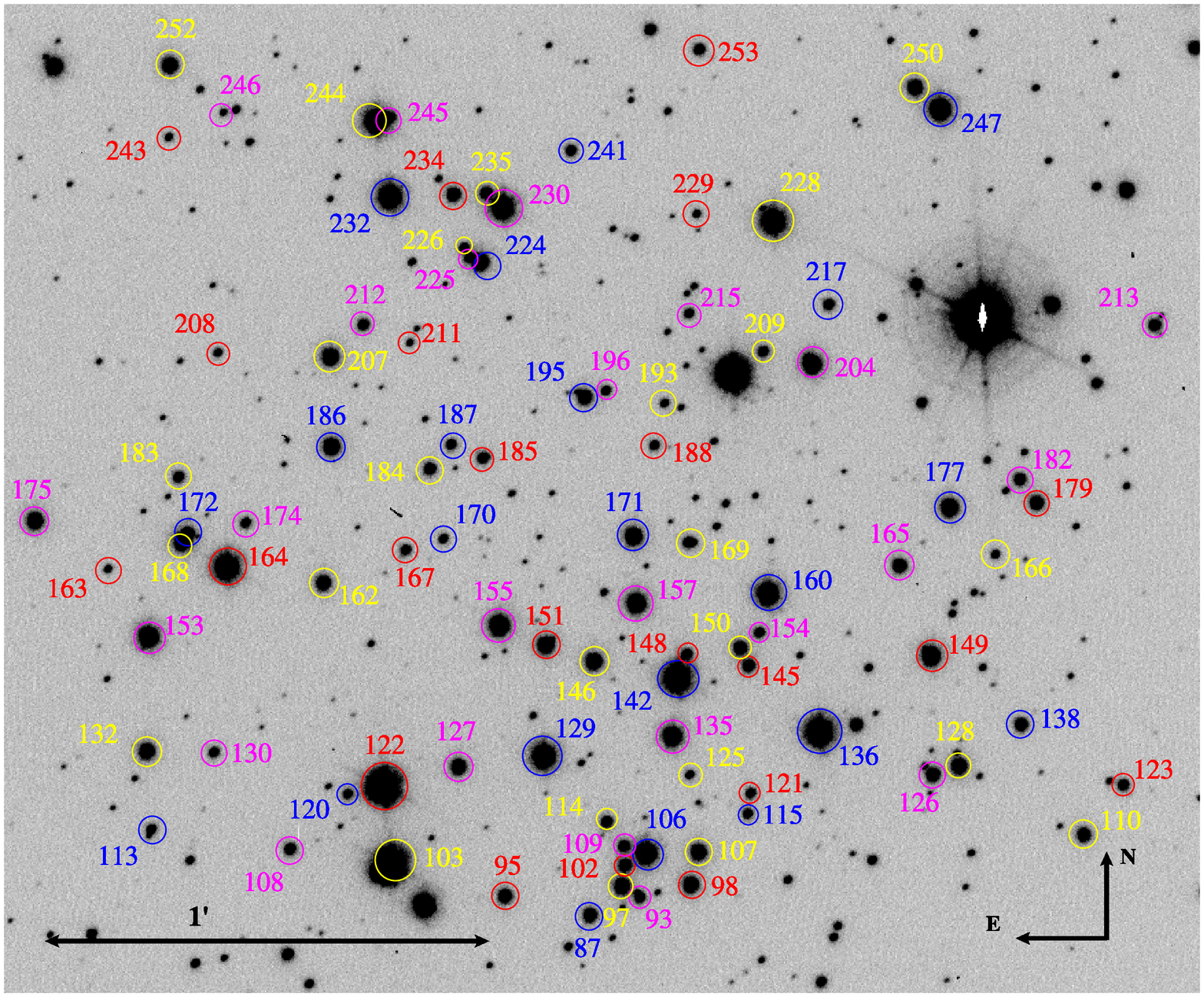}}
\centering
\caption{Finding chart for stars with photometry in the central part of NGC~7419. The image is one of our $V$-band frames. Each star is identified by the nearest marker in the same colour as the circle around it. The origin of the coordinates is located at the bottom-left corner of the frame.}
\label{Fig2} 
\end{figure*}

The number of stars that we could detect in all filters is limited by
the long exposure in the $u$ filter.
We identify all stars with good photometry photometric errors $\le 0.04$ in all four filters on the image in Figures~\ref{Fig1} and~\ref{Fig2}. In Table~\ref{tb4}, we list their coordinates in J2000 \citep[from][]{joshi2008} and their identification with objects in the WEBDA database. The designation of each star is given by the number indicated on the images (Fig.~\ref{Fig1} and Fig.~\ref{Fig2}). We have photometry for 202 stars in the field. In Table~\ref{tb4}, we list the values of $V$, $(b-y)$ and 
$c_{1}$ with the assigned photometric errors and the number of measurements for each magnitude, colour, or index. The error assigned is the standard deviation for star with three or four measurements and the photometric errors from DAOPHOT for stars with only one measurement, with the errors in the colour indices calculated through error propagation. 

We completed our dataset with $JHK_{s}$ photometry from the 2MASS
catalogue \citep{Skrutskie2006}. In Table~\ref{irtab} we show the 2MASS photometric values for the
stars confirmed as cluster members in Section~\ref{HR} that have good-quality photometric flags and photometric errors below 0.05 mag in all 2MASS filters.

\subsection{Spectroscopy}

Most of the spectroscopic observations were obtained with ALFOSC in spectroscopic mode during the 2004 and 2005 NOT runs. A few spectra of selected stars were taken in service mode in November 2010 (on the nights of November 10 and November 23). During the 2004 run, the seeing was quite poor, except on the night of November 3. We resorted to using a $1\farcs8$ slit on both November 2 and 4. For all other ALFOSC observations, we used $1\farcs0$ slits.

Our primary setup was grism \#16 for classification spectroscopy. This grism covers the 3500--5060 \AA\ range with a nominal dispersion of 0.8 \AA/pixel. The resolving power, measured on arc lines, is $\sim1\,500$ with the $1\farcs0$ slit. With the $1\farcs8$ slit, it even reaches $R\sim1\,000$.

In the 2004 run, a few objects were observed with grism \#7, which covers the 3850--6850\AA\ with a nominal dispersion of 1.5\AA/pixel. The resolving power is $\sim 550$ with the $1\farcs8$ slit and $\sim700$ with the $1\farcs0$ slit. In the 2005 run, we observed a few fainter stars with grism \#14. This grism covers the 3275--6125\AA\ range with a nominal dispersion of 1.4\AA/pixel. The resolving power with the $1\farcs0$ slit is $R\sim800$. Even though this grism does not include the region of H$\alpha$, it has a much higher throughput in the blue than grism \#7, and it was thus preferred to obtain classification spectra for faint stars. Finally, in the 2010 service observations, two objects were observed with grism \#8. This grism covers the 5825--8350\AA\ range with a nominal dispersion of 1.3\AA/pixel, giving a resolving power of $R\approx1300$. Wavelength calibration was achieved by using ThAr arc lamp spectra, taken between the exposures. The rms for the wavelength solution is $\approx0.2$ pixels for all grisms.

During the 2004 run, we also obtained slitless spectroscopy of the field by combining the low-resolution
grism \#4 with the Bessell $R$-band filter. Slitless fields were taken at three different orientations
to minimise source overlapping. More slitless observations were taken during other runs in August 2008 and September 2011.

Low-resolution spectra of some of the brightest Be candidates were taken on 2001 July 1 with the 1.93-m telescope at
Haute Provence Observatory (France) and the Carelec spectrograph, equipped with an EEV42 CCD. A  $1\farcs8$ slit was used, giving a resolving power $R\approx1000$ over the 3800--6800 range, though the signal-to-noise ratio (S/N) is very low in the blue.

\begin{figure}
\centering
\resizebox{\columnwidth}{!}{\includegraphics[angle=-90]{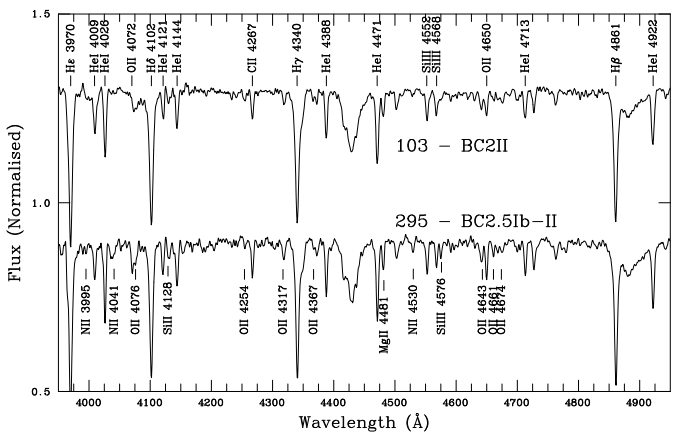}}
\caption{Classification spectra of the two brightest blue cluster members obtained with the NOT. Note the lower resolution of the top spectrum, observed with a $1\farcs8$ slit.The most prominent stellar features are marked. The strong diffuse interstellar bands at 4428, 4502, 4726, 4760, and 4882\AA\ are not marked.}
\label{fig:bsgs}
\end{figure}

Spectra of the red supergiants in the far-red were taken on 2007 August 21 using the red arm of the ISIS double-beam spectrograph, mounted on the 4.2 m William Herschel Telescope (WHT) in La Palma (Spain). The instrument was fitted with the R600R grating and the Red+ CCD. This configuration covers the 7600-9000\AA\ range in the unvignetted section of the CCD with a nominal
dispersion of 0.5\AA\/pixel. The CCD was binned by a factor of two in the spectral direction and a $1\farcs2$ slit was used. The resolution element is expected to be about four unbinned pixels. This was checked by measuring the width
of arc lines, which is on average $\approx2.1\AA$. The resolving power of our spectra is $R\approx4000$.
The observation of the star B139 was taken on 2009 September 13 with the same configuration and a $1\farcs5$ slit. 

All spectroscopic data were reduced using the Starlink software packages CCDPACK \citep{Draper2000} and FIGARO \citep{Shortridge1997}. We used standard procedures for bias subtraction and flat-fielding (using internal halogen lamps). The spectra were normalised to the continuum using DIPSO \citep{Howarth1998}.

\section{Results}

\subsection{Spectroscopy}

In total, we obtained blue spectra for $\sim$ 40 stars, but not all of them are useful for spectral classification. For the objects with spectra of sufficient quality for an accurate spectral classification, spectral types are given in Table~\ref{types}.

\subsubsection{Spectral classification}
\label{class}
The brightest blue stars in the cluster are star 103, in the core region, and star 295, an outlier to the northwest. Their spectra, displayed in Fig.~\ref{fig:bsgs}, are very similar. They are rather luminous stars. The spectral types of early-B stars are commonly determined by comparing ratios of different Si ions. In these two stars, \ion{Si}{iv} is undetectable, while \ion{Si}{iii} lines are strong and \ion{Si}{ii} lines are moderately strong. This places our objects close to the B2 spectral type \citep{walborn1990}. The \ion{Si}{ii}~4128--30\AA\ doublet is too weak compared to the neighbouring \ion{He}{i} lines for a B3 spectral type, while the \ion{O}{ii} spectrum is too weak for B1. The main luminosity criterion around B2 is the ratio of \ion{Si}{iii}~4552\AA\ to \ion{H}{i}~4387\AA\  \citep{walborn1990}. In our objects, it indicates a luminosity class of II or Ib. The very strong \ion{C}{ii}~4267\AA\ line could suggest that the spectral type is B2.5, also supported by the relatively high ratio of \ion{Mg}{ii}~4481\AA\ to the \ion{Si}{iii} lines \citep{walborn1990}. However, at this spectral type the \ion{N}{ii} spectrum should be close to its maximum, but N lines are very weak or absent in our targets. \ion{N}{ii}~3995\AA\ is very weak compared to \ion{He}{i}~4009\AA, while \ion{N}{ii}~4631\AA\ is absent, when it should be stronger than \ion{O}{ii}~4650\AA. This strongly indicates that both stars are nitrogen-deficient. If this is the case, they are likely to be carbon-enhanced as well. We assign nominal spectral types BC2\,II for star 103 and BC2.5\,Ib-II for star 295, but note that with the uncertainties discussed, both stars could be given identical classifications, within the B2--B2.5 range and a luminosity class very slightly lower than necessary to be a supergiant.

\begin{figure}
\centering
\resizebox{\columnwidth}{!}{\includegraphics{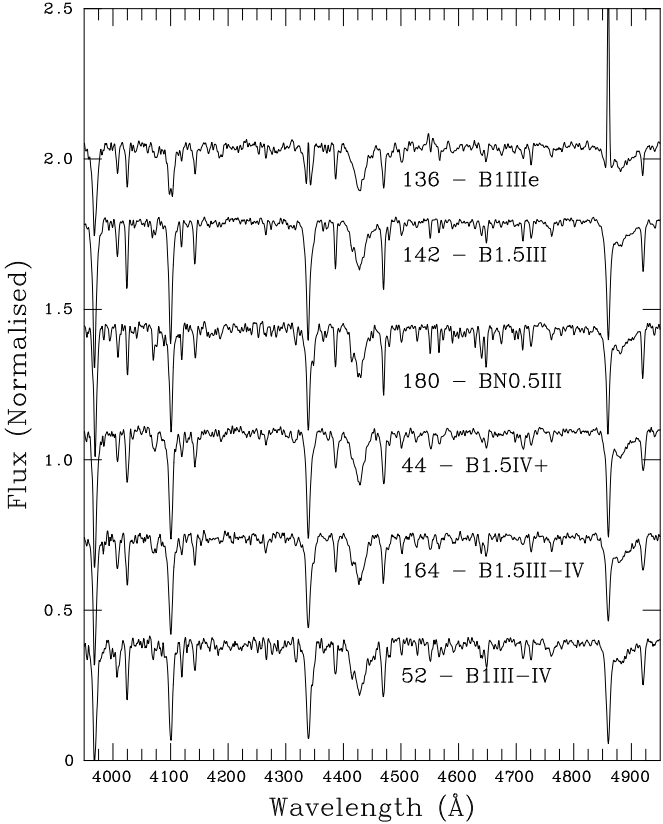}}
\caption{Classification spectra of several bright blue cluster members.
\label{fig:top2}}
\end{figure}

The spectra of several other bright cluster members are shown in Fig.~\ref{fig:top2}. The next-brightest blue star is star 180. This object is clearly earlier than any other member, as shown by the presence of \ion{Si}iv~4089\AA, or the prominent \ion{C}{iii}+\ion{O}{ii}~4650\AA, which is as strong as \ion{H}{i}~4387\AA. Its position in the photometric colour-magnitude diagram (CMD) (see Fig.~\ref{vc1}) confirms that this object is a blue straggler. The prominent \ion{C}{iii}+\ion{O}{ii}~4070\AA, as deep as  \ion{He}~4144\AA, points to a spectral type as early as B0.5\,III, in good agreement with the $c_{0}$ value, even if \ion{He}{ii}~4686\AA\ is not seen at this resolution. However, again, there is evidence for some anomaly, as nitrogen lines should not be seen at this spectral type. \ion{N}{ii}~3995, 4041, 4530, and 4631\AA\ are present, though, as well as several weak lines between 4601--4621\AA. Therefore the star is likely to be slightly nitrogen-enhanced. Nitrogen enhancement is usually accompanied by carbon deficiency, which does not seem to be present. In view of this, we classify the star as B0.5N\,III, though we cannot rule out the possibility that it is slightly later and more luminous, around B1\,II.

The brightest Be star is star 136, which presents heavy veiling by emission lines. Its spectral type is close to B1\,III. It seems to be slightly less luminous, but this may be due to the broad shallow lines typical of Be stars. Star 52 has a similar spectral type, but does not present emission lines. Star 142 is slightly later, as shown by the stronger \ion{C}{ii}~4267\AA\ line and the presence of some weak \ion{N}{ii} lines. We assign a spectral type B1.5\,III. The Be star 164 has a similar spectral type. This later star showed emission lines until 2005, but presents a completely normal spectrum in 2010 (displayed).

Star 44 is an interesting case. We have three spectra of this object (2004, 2005, and 2010), all looking different. The best one, from 2010 (displayed), shows a normal B1.5\,IV star. This spectral type does not seem to agree with its brightness. Moreover, the 2004 spectrum, though rather noisier, showed evidence for a later spectrum, with strong \ion{Si}{ii}~4128--30\AA\ and \ion{C}{ii}~4267\AA\ and weak \ion{Si}{iii}. We speculate that the star is perhaps a binary system, containing a B1.5\,IV and a $\sim$B3\,III star, though more observations are needed to confirm this. The star displayed some emission infilling in H$\alpha$ in the 2001 spectrum, but no evidence for Be characteristics in any of the blue spectra. \citet{caron2003} classified this star as B4\,III based on its far-red spectrum, corroborating the presence of a late-type companion. They remarked that the Paschen lines were flat-bottomed, an effect that they attributed to weak Be characteristics, but that could also be due to the presence of two stellar spectra. \citet{malchenko2011} observed weak double-peaked H$\alpha$ in this object in June 2006.

\setcounter{table}{4}
The spectra of fainter stars have a lower resolution or a lower S/N (or both), but their features are still good enough for rough classifications, accurate to $\pm1$ spectral type. There is a number of relatively bright Be stars (129, 133, 228, 230, 232, and 244) that we classify as B1--1.5\,IV. These objects represent the top of the main sequence, and the concentration of Be stars seems to be very high in this magnitude range. Around $V=15$, we find a number of stars without emission lines (e.g., 106, 160) that are definitely not giants. These objects with spectral type $\approx$B1\,IV indicate the turn-off age of the cluster. Unfortunately, the quality of the spectra does not allow us to give high weight to the luminosity-class determination. Star 247, on the other hand, shows a very weak \ion{O}{ii} spectrum, and it may be a more evolved star, B2\,III-IV, its faintness explained by higher-than-average reddening. Unfortunately, its spectrum has a very low S/N. 

\begin{table}
\caption{Spectral types for stars in NGC~7419 with classification spectra.\label{types}}
\centering
\begin{tabular}{lc}
\hline\hline
\noalign{\smallskip}
Star&Spectral type\\
\hline
\noalign{\smallskip}
\multicolumn{2}{c}{Red supergiants}\\
\hline	
\noalign{\smallskip}
B139	&	M1\,Iab		\\
B435	&	M1.5\,Iab       \\
B696=122&	M1.5\,Iab       \\						
B921=56	&	M0\,Iab		\\					  
B950	&	M7.5\,I		\\					  
\hline	
\noalign{\smallskip}		  
\multicolumn{2}{c}{Be stars}			\\
\noalign{\smallskip}	
\hline	
\noalign{\smallskip}						 
129	&	B1.5\,IVe	\\					  
133	&	B1.5\,IVe	\\					  
136	&	B1\,IIIe	\\					  
155	&	B1\,Ve		\\
164	&	B1.5\,III-IVe	\\				  
207	&	B2\,Ve		\\					  
228	&	B1\,IVe		\\					  
230	&	B1.5\,IVe	\\					  
232	&	B1\,IVe		\\					  
244	&	B1\,IVe		\\
\hline
\noalign{\smallskip}	
\multicolumn{2}{c}{Other bright blue stars}		\\
\noalign{\smallskip}	
\hline
\noalign{\smallskip}		
44	&	B1.5\,IV+?\,(e)\\			
52	&	B1\,III-IV	\\
91	&	B1\,V		\\	
103	&	BC2\,II		\\
106	&	B1\,IV		\\
142	&	B1.5\,III	\\	
160	&	B1.5\,IV	\\
180	&	BN0.5\,III	\\	
247	&	B2\,III-IV	\\					  
295	&	BC2.5\,Ib-II	\\
\hline						
\end{tabular}
\end{table}

A peculiar case is star 91. This is the most reddened B-type star in the cluster. Even though it is quite faint in the $B$ band, its dereddened magnitudes indicate $M_{V}=-4.9$, far too bright for a main-sequence star. We only have a low-resolution spectrum of this object, which suggests a spectral type B1\,V. Despite the low resolution, we should expect to see some metallic lines in the spectrum if the object were a B1\,III star, as suggested by its absolute  magnitude, since the S/N is good (60--80 in the classification region). Perhaps the star is a background object in the Outer or Cygnus arm, but a better spectrum is needed to ascertain this possibility.

\subsubsection{Red supergiants}
\label{RSGs}

The presence of five red supergiants in NGC~7419 was first noted by \citet{blanco}. Using objective-prism infrared spectra, they derived spectral types between M2 and M3.5 for four of them, and M7 for the last one. This later-type star was observed to be photometrically variable. \citet{fc74} took near-infrared photographic spectra of four of the red supergiants, confirming that they all had the same radial velocity. They classified three of them as M2\,Iab (B921, B696, and B435)\footnote{For the red supergiants, we use the notation of \citet{beauchamp1994}, because only two of them were detected in all filters by our photometry, B696=122 and B921=56.}, and confirmed the late spectral type of B950, which they refined to M7.5\,I. Finally, \citet{beauchamp1994} obtained CCD spectra of all five RSGs in the region of the \ion{Ca}{ii} triplet, confirming that they have radial velocities very similar to blue cluster members. They also confirmed their RSG character by measuring the strength of the \ion{Ca}{ii} triplet lines.

\begin{figure}
\centering
\resizebox{\columnwidth}{!}{\includegraphics{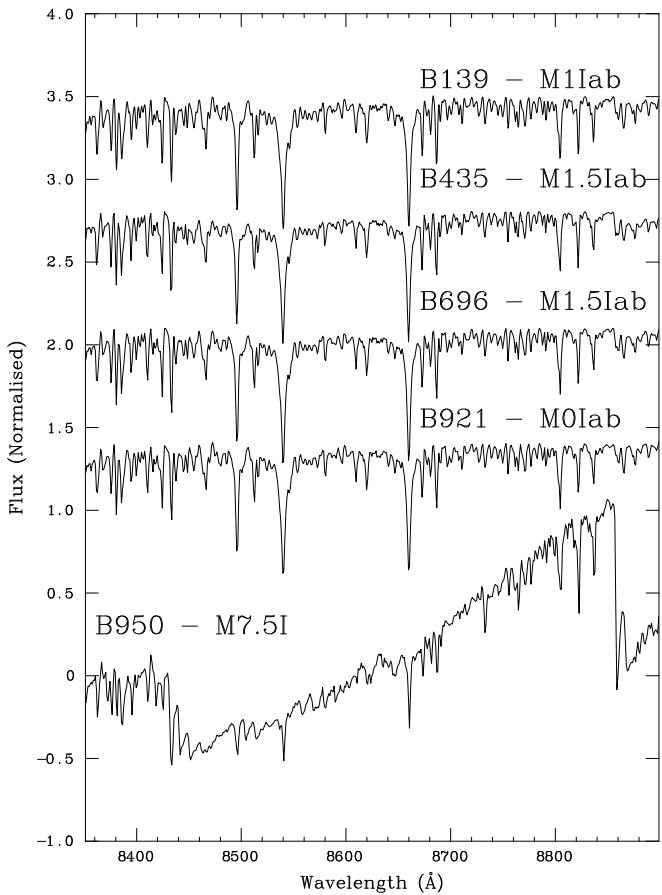}}
\caption{Spectra of the five red supergiants in
NGC 7419, in the region around the \ion{Ca}{ii} triplet.
\label{msgs}}
\end{figure}

Figure~\ref{msgs} displays the spectra of the five RSGs around the \ion{Ca}{ii} triplet. The criteria for spectral classification of RSGs in this regions have been summarised by \citet{negueruela11}. In this range, the spectral type can be determined from the depth of the TiO $\delta$(0,0) $R$-branch bandhead at 8860\AA, which correlates strongly with effective temperature ($T_{{\rm eff}}$). B921 has no sign of this bandhead and therefore is around M0. B139, B435, and B696 have weak bandheads that make them earlier than M2. The luminosity class can be estimated from the \ion{Ti}{i}/\ion{Fe}{i}~8468\AA\ blend and the equivalent width of the \ion{Ca}{ii} lines, which suggest class Iab for all of them. The high luminosity is also supported by the very high ratios of the 8514\AA\ blend to the \ion{Ti}{i}~8518\AA\ line \citep{kh45}. However, the blend is still rather weaker than \ion{Ti}{i}~8435\AA, showing that the luminosity class is not Ia.

 B950 is much later, with very strong TiO systems.
Classification criteria for RSGs later than M5 have not been described. Standard stars have not been defined, because late-M supergiants generally show spectral-type variability \citep{abt63}. \citet{solf} extended classification criteria to late-M giants in the red region.
According to these, at M7, the VO bandhead at 7982\AA\ is of similar strength to the TiO 7820, 7828\AA\ banheads, while the VO feature at 7961\AA\ is just visible. The VO bandheads at 8538 and 8624\AA\ become first visible at M8 on photographic spectra, with the former being comparable in strength to \ion{Ca}{ii}~8542\AA. Our spectrum of B950 shows much stronger \ion{Ca}{ii}, a fact attributable to the dependence on luminosity of the \ion{Ca}{ii} triplet. The 8624\AA\ bandhead is clearly seen, though blended with the 8621\AA\ interstellar band. The 7961\AA\ VO bandhead is also quite strong. In view of this, and allowing for the differences between giants and supergiants, we find the M7.5\,I classification fully justified. The resulting classifications are summarised in Table~\ref{types}.

The classifications for B435, B696, and B950 are fully compatible with previous work. However, B139 and B921 are earlier here. B921 is clearly earlier than any other RSG in the cluster, and given as M0\,Iab, against M2.5\,I in \citet{blanco} and M2\,Iab in \citet{fc74}. B139 is M1\,Iab, against M3.5\,I in \citet{blanco}. Even though the differences of two spectral subtypes seem large, classification criteria are subtle and the weight of prism spectral types is low. Since most early-M supergiants show stable spectral types, additional observations are needed before spectral variability can be claimed.

\subsubsection{Be stars}
\label{bes}

Previous authors have reported a high percentage of Be stars in NGC~7419 \citep{pigulski2000, subramanian2006, joshi2008}. All these references reported single-epoch observations. Be characteristics, on the other hand, are known to be variable on time scales ranging from months to many years \citep{porter2003}. For the first time, we present multi-epoch spectroscopy of Be stars in NGC~7419, which we complement with all the data available in the literature. Table~\ref{Bevar} lists all members of NGC~7419 that have ever been reported to show emission in H$\alpha$, together with all detections of emission for every star. For each observation or each reference in the literature, we indicate detections, reported non-detections, or lack of definite evidence (which may have many different causes, such as crowding in slitless fields, location outside the field of view of the observations, or lack of the information in the corresponding references). In November 2010, four stars listed in Table~\ref{Bevar} were observed in service mode. Classification spectra were obtained for 44 and 164, and showed no sign of emission (shown in Fig.~\ref{fig:top2}). Spectra around H$\alpha$ were taken for 151 and 155 (shown in Fig.~\ref{fig:alphas}). Since only four stars were observed, this run is not reported in Table~\ref{Bevar}.    
 
\begin{table}
\caption{Spectroscopic monitoring of Be stars for ten years. The ''X'' symbol means that we cannot know if the source was displaying emission line because it was outside the field, heavily blended, too faint for detection, or some other reason. ''$-$'' means that the star did not show emission in H$\alpha$. ``+'' means that the star displayed the H$\alpha$ line in emission.\label{Bevar}}
\begin{center}
\begin{tabular}{lcccccccccc}
\hline\hline
\noalign{\smallskip}
\noalign{\smallskip}
Name&A\tablefootmark{1}&B\tablefootmark{2}&C\tablefootmark{3}&D\tablefootmark{4}&E\tablefootmark{5}&F\tablefootmark{6}&G\tablefootmark{7}&H\tablefootmark{8}&I\tablefootmark{9}&K\tablefootmark{10}\\	
\hline
\noalign{\smallskip}		
18&X&	X&X&+&X	&$-$&X&+&+&	+\\
30&+&	X&X&X&X	&+&X&+&+&	$-$\\			  
34&$-$&	X&X&X&X	&$-$&X&$-$&$-$&	+\\
44&+&	+&X&X&$-$	&$-$&+&+&$-$&	$-$\\				
47&+&	X&X&+&+	&+&+&+&+&	+\\			
55&+&	X&X&+&X	&+&+&+&+&	+\\
61&+&	X&X&X&X	&$-$&$-$&+&$-$&	$-$\\				
62&+&	X&X&+&X	&+&+&+&+&	+\\
87&+&	X&X&X&X	&$-$&X&+&+&	$-$\\				
96&+&	X&+&+&X	&+&+&+&+&	+\\	
98&+&	X&X&X&X	&$-$&$-$&+&+&	$-$\\			
100&+&	X&+&X&X	&X&X&X&+&	+\\
107&+&	X&X&X&X	&$-$&+&+&$-$&	$-$\\	
126&+&	+&+&X&X	&$-$&X&+&+&	$-$\\					
129&+&	X&+&+&+	&+&+&+&+&	+\\			
133&+&	+&X&+&+	&+&+&+&+&	+\\			
136&+&	+&+&+&+	&+&X&+&+&	+\\
151&+&	+&+&X&X	&+&X&+&$-$&	$-$\\				
153&+&	X&X&+&+	&+&+&+&+&	+\\			
155&+&	+&$-$&$-$&$-$	&+&+&+&$-$&	+\\			
157&+&	X&X&+&X	&+&+&+&$-$&	+\\
164&+&	+&+&+&+	&+&+&+&$-$&	$-$\\				
168&+&	+&X&+&X	&+&X&+&+&	+\\
171&$-$&	X&X&X&X	&$-$&X&$-$&+&	$-$\\				
175&+&	+&X&+&X	&+&+&+&+&	+\\			
207&+&	X&X&+&+	&+&+&+&+&	+\\			   
228&+&	+&+&+&+	&+&+&+&+&	+\\			   
230&+&	+&+&+&+	&+&+&+&+&	+\\			   
232&+&	+&+&+&+	&+&+&+&+&	+\\			   
235&+&	X&X&$-$&X	&X&X&X&+&	+\\			   
244&+&	+&+&+&X	&+&+&+&+&	+\\			   
250&+&	X&X&X&X	&X&+&+&+&	+\\			   
258&+&	X&X&+&X	&+&+&+&+&	+\\
260&$-$&	X&X&X&X	&$-$&X&+&+&	$-$\\	
273&X&	X&X&X&X	&$-$&X&$-$&+&	$-$\\			   
281&X&	X&X&X&X	&$-$&X&$-$&X&	+\\			   
290&X&	X&X&X&X	&$-$&X&+&+&	+\\			   
294&X&	X&X&X&X	&$-$&X&+&X&	+\\			  		  	\hline
\end{tabular}
\newline\\
\tablefoottext{1}{\citep{pigulski2000}},
\tablefoottext{2}{2001(spectra)},
\tablefoottext{3}{2004(spectra)},
\tablefoottext{4}{2004(slitless)},
\tablefoottext{5}{2005(spectra)},
\tablefoottext{6}{\citep{subramanian2006}},
\tablefoottext{7}{\citep{malchenko2011}},
\tablefoottext{9}{\citep{joshi2008}},
\tablefoottext{8}{2008(slitless)},
\tablefoottext{10}{2011(slitless)}
\end{center}
\end{table}

Using the standard definition of a Be star as a star that has or has had Balmer lines in emission at one time \citep{porter2003}, we consider any star ever reported as showing emission as a Be star. As can be seen in Figures~\ref{vc1}, \ref{Fig3}, and~\ref{Fig5}, Be stars concentrate above and around the main-sequence turn-off, a behaviour observed in many other clusters \citep{mcSwain2005}. The fraction of Be stars to total number (B+Be) is $58\%$ for stars brighter than $V=15$ (essentially, above the turn-off) and $40\%$ for stars brighter than $V=17$ (corresponding to spectral type B2\,V). Below this limit the number of Be stars is small. Though selection effects might play a role in this low fraction of Be stars at faint magnitudes, we consider this unlikely, because different authors have used H$\alpha$ photometry, which is sensitive at faint magnitudes. For example, \citet{joshi2008} claimed to be essentially complete down to $V=18$ and have a completeness factor of 90\% in the cluster halo down to $V=20$. We must thus conclude that the fraction of Be stars decreases as we move down the main sequence towards late-B types, as found in most clusters containing classical Be stars \citep{mcSwain2005}.  

\begin{figure}
\centering
\resizebox{\columnwidth}{!}{\includegraphics[angle=-90,bb=70 110 500 700,clip]{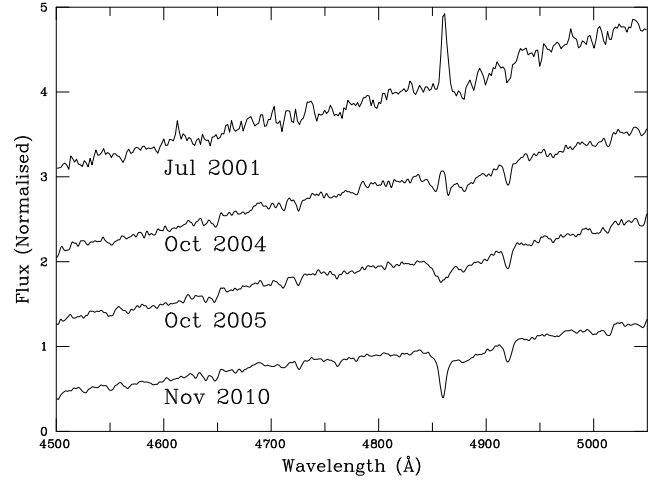}}
\caption{Spectra of the Be star 164 taken at different epochs and showing the progressive disappearance
of emission characteristics.\label{fig:varia2}}
\end{figure}

\begin{figure}
\centering
\resizebox{\columnwidth}{!}{\includegraphics[angle=-90,bb=60 115 500 723,clip]{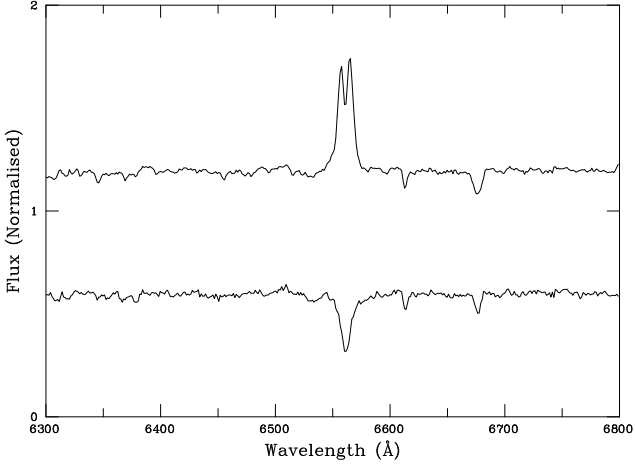}}
\caption{Intermediate-resolution spectra around H$\alpha$ for 155 (top) and 151 (bottom) in November 2010. These two Be stars have undergone substantial variability during the past ten years, including the complete loss of the envelope. At the time when the spectrum was taken, 155 was reforming its disk.\label{fig:alphas}}
\end{figure}

The information displayed in Table~\ref{Bevar} fully confirms the variable nature of Be characteristics in many of the stars observed. For example, in 2001, star 164 had H$\alpha$ strongly in emission ($EW=-28$\AA), while H$\beta$ was in emission well above the continuum. Its subsequent evolution can be seen in Fig.~\ref{fig:varia2}. The emission line gradually disappeared between 2001 and 2005, until the spectrum taken in 2010 (displayed in Fig.~\ref{fig:top2}) shows no characteristics that allow its classification as a Be star. H$\alpha$ was still weakly in emission in 2006 \citep{malchenko2011}. A similar evolution was followed by Star 151. This object displayed strong H$\alpha$ emission ($EW=-25$\AA) in 2001, but this had decreased strongly by 2004 ($EW=-13$\AA). Emission from this star was not detected in the slitless spectroscopy either in 2008 and 2011, while the spectrum taken in 2010 (Fig.~\ref{fig:alphas}) shows the H$\alpha$ line fully in absorption and lacking any emission component.

More complex evolution has been shown by star 155. Detected as a Be star by \citet{pigulski2000}, it displayed a neutralised H$\alpha$ line (the emission component is strong enough just to mask the photospheric absorption) in its 2001 spectrum. After this, it was not detected in emission in 2004 or 2005, but reappeared as an emission-line star in the 2008 slitless spectrum. \citet{malchenko2011} had already detected a very weak double-peaked emission in 2006. In November 2010, it displayed a moderately strong double-peaked emission line (Fig.~\ref{fig:alphas}), indicative of a circumstellar disk well in the process of reformation.

This sort of variability is typical of classical Be stars, but is not observed in Herbig Be stars. Table~\ref{Bevar} shows many other examples of stars that appear to have passed through diskless phases, though they are not as well documented as the examples above. The time-scales implied are typical of classical Be stars with disk disappearance and reformation happening over a few years.

\subsection{Observational HR diagram}
\label{HR}

\begin{figure}
\resizebox{\columnwidth}{!}{\includegraphics[clip]{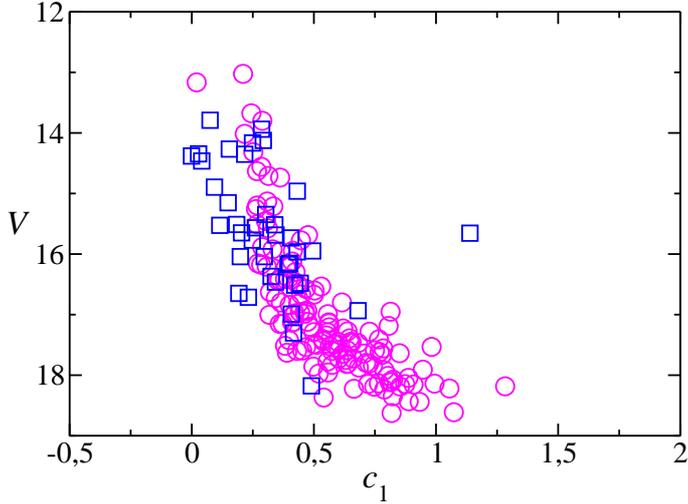}}
\caption{$V/c_{1}$ diagram for all members in the cluster. Open squares represent the members catalogued as Be stars.\label{vc1}}
\end{figure}

\begin{figure}
   \resizebox{\columnwidth}{!}{\includegraphics[clip]{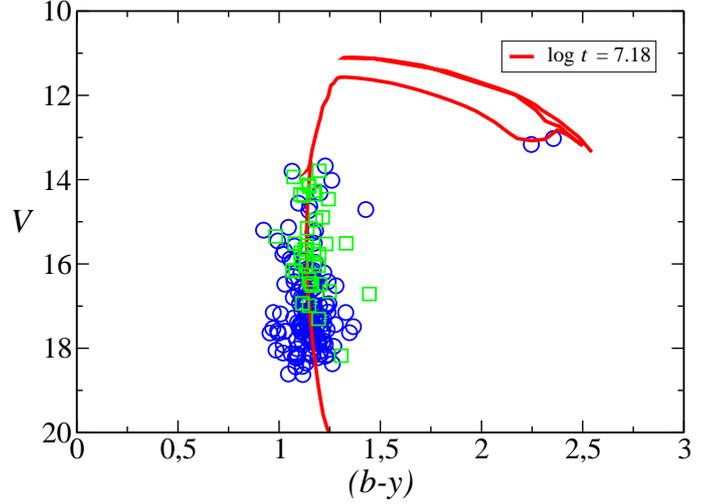}}
   \caption{$V/(b-y)$ diagram for all members in the cluster. Open squares represent the members catalogued as Be stars. The solid line is the isochrone of \citet{marigo2008} corresponding to $\log t = 7.18$, reddened by $A_{V}=5.2$ and displaced to $DM=13.0$. The isochrone has been reddened taking into account colour effects, following the procedure described in \citet{girardi2008}.\label{Fig3}}
    \end{figure}

\begin{figure}
   \centering
\resizebox{\columnwidth}{!}{\includegraphics[clip]{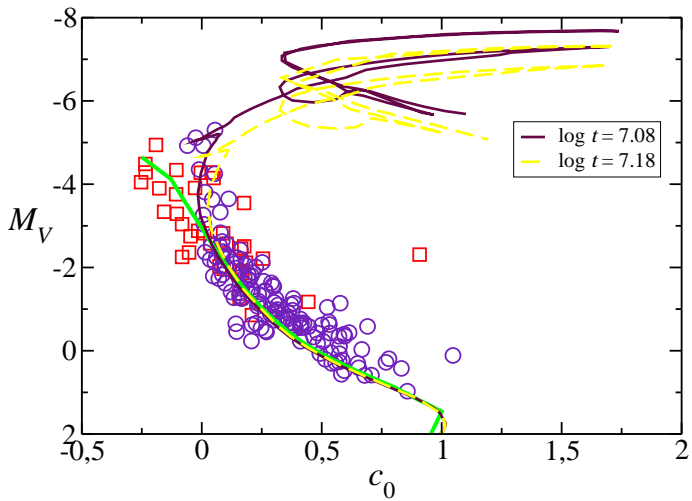}}
\caption{Absolute magnitude $M_{V}$ against intrinsic colour $c_{0}$ for B-type stars in NGC~7419. Open squares are stars catalogued as Be stars. The thick line represents the ZAMS from \citet{perry1987}. Two isochrones of \citet{marigo2008}, corresponding to $\log t = 7.08$ and $\log t = 7.18,$ are labelled with their respective $\log t$.\label{Fig5}}
 \end{figure}

We started the photometric analysis by plotting the $V/(b-y)$ and $V/c_{1}$ diagrams for all stars in our field. To assess the membership of a star, we look at its position in the $V/(b-y)$ and $V/c_{1}$ diagrams. We found that in both diagrams, the vast majority of the stars fall along a very well defined main sequence. We considered as non-members the objects that do not fit the main-sequence loci in either of the diagrams ell, unless they are catalogued as Be stars. Indeed, Be stars have colours different from those of non-emission B stars because of the additional reddening that is caused by the circumstellar envelope, and they tend to have redder $(b-y)$ and lower $c_{1}$ values
than normal B stars (see Figures~\ref{vc1} and~\ref{Fig3}).

We calculated individual
reddenings for all cluster members. We followed the procedure described
by \citet{crawford1970b}, and we used the observed $c_{1}$ to
predict the first approximation to $(b-y)_{0}$ with the expression
$(b−-y)_{0}$ = $−-0.116$ + $0.097$$c_{1}$. Then we calculated
$E(b−-y)$ = $(b−-y)$$-−$$(b-y)$$_{0}$ and used $E(c_{1})$ = $0.2$$E(b−-y)$
to correct $c_{1}$ for reddening $c_{0}$ = $c_{1}$$-−$$E(c_{1})$. The intrinsic
colour $(b−-y)_{0}$ was then calculated by replacing $c_{1}$ with $c_{0}$  in the
above equation for $(b-y)_{0}$ . Three iterations are enough to
reach convergence in the process. Naturally, this procedure only results in physically meaningful values for stars without circumstellar excesses (i.e., not for Be stars). Restricting the analysis to stars that have never been classified as Be stars, we found a moderate degree of differential reddening amongst members. The average value
of the colour excess is $E(b−-y)$ = 1.23$\pm0.12$, where the error bars are given by the standard deviations of all values\footnote{All colour excesses are within two standard deviations of the average, with the single exception of star 91, showing that the spread in reddening is moderate (see Fig.~\ref{Fig7}).}. The median value is $E(b-y) = 1.24$.

Using the standard relation $E(b-y)=0.74E(B-V)$ and taking into account the presence of differential reddening, this average value agrees very well with the $E(B-V)\approx1.7$ determined by \citet{joshi2008} and \citet{subramanian2006}\footnote{The agreement in the $E(B-V)$ determination between these two references is surprising, since large systematic differences exist between their $(U-B)$ values.}. Our value disagrees from that
found by \citet{beauchamp1994}, $E(B-V)\approx2.0$   even though we found the same age for the cluster (see Section~\ref{age}). This confirms the suspicion by \citet{beauchamp1994} that their $(U-B)$ colours, calibrated against stars in NGC~7510, might suffer from a systematic error. 

With the aid of these individual values we calculated 
the intrinsic colours $c_{0}$ and magnitudes $V_{0}$ of the normal B-type members. From $c_{0}$ and $V_{0}$, we estimated photometric spectral types using the calibration for $c_{0}$ from \citet{crawford1978}. The photometric types are displayed in Table~\ref{photparam}. For those stars with spectroscopic spectral types (see Table~\ref{types}), the agreement between photometric and spectroscopic types is excellent, certifying the correctness of the dereddening procedure.

\subsubsection{Distance determination}
\label{distance}

We estimated the distance modulus ($DM$) to NGC~7419
by fitting the observed $V_{0}$ vs. $c_{0}$ zero age main-sequence (ZAMS) to the mean calibrations of
\citet{perry1987}. We fitted the ZAMS as a lower envelope for most of the members, deriving a best-fit distance modulus of $V_{0}−-M_{V}$ = $13.0\pm0.2$ (the error indicates the uncertainty in positioning the theoretical
ZAMS and its identification as a lower envelope; see Fig~\ref{Fig5}). This $DM$ corresponds to a distance of $4.0^{+0.4}_{-0.4}$ kpc. Both \citet{joshi2008} and \citet{subramanian2006} found distances around $\approx 3$ kpc, but their error bars place our value within 2$\sigma$ of our determination. Our data were individually dereddened while theirs are based on average reddening, but the difference in $DM$ results almost exclusively from the different ages found (see next subsection). \citet{subramanian2006} and \citet{joshi2008} used a fit to theoretical isochrones to simultaneously determine the age and distance to the cluster, while we fitted the distance to the observational ZAMS.  
The difference with respect to the value of \citet{beauchamp1994} is due to the adoption of a different reddening law.

\subsubsection{Age determination}
\label{age}

To determine the cluster age, we fitted by eye isochrones in the $V/(b-y)$ (Figure~\ref{Fig3}) and $M_{V}/c_{0}$ (Figure~\ref{Fig5}) observational diagrams to the position of the red stars and the turn-off. We used isochrones from \citet{marigo2008}, computed using a \citet{kroupa1998} initial mass function (IMF) corrected for binaries. The isochrone shown in Fig.~\ref{Fig3} was reddened following the procedure described in \citet{girardi2008}, which takes into account colour effects for late-type stars. The reddening procedure uses the extinction law from \citet{cardelli1989} and \citet{odonnell1994} with $R_{V}=3.1$.

The $M_{V}/c_{0}$ diagram provides a good fit of the turn-off for the 12~Myr isochrone, once the Be stars are removed. The brightest B stars are close to the end of H-core burning, which agrees very well with their observed spectral types. The turn-off can also be fitted with a slightly older isochrone (15~Myr). The brightest blue stars then appear as blue stragglers, but the fit to the turn-off improves slightly. Spectroscopically, only star 180 (BN0.5\,III) is decidedly a blue straggler. Since the dereddening method used to derive $c_{0}$ is only valid for early-type stars, the red supergiants 56 and 122 do not appear in this diagram.

The two red supergiants appear in the $V/(b-y)$ diagram, their position agrees quite well with the 15~Myr isochrone, though their location is not exact. The 12~Myr isochrone only gives a poor fit for these stars, though again both isochrones fit the blue sequence well. We cannot place too much weight on the position of the red supergiants, because the reddening to the cluster is very high, and the polynomial form used by \citet{cardelli1989} introduces artificial bumpiness in the extinction curve \citep{maiz2012}. 

In view of this, we take the cluster age to be $\tau = 14\pm2$~Myr, a range covering all isochrones that give a good fit to the dereddened $M_{V}/c_{0}$ plot. This age is almost identical to that found for the double cluster h and $\chi$~Per \citep{currie2010}, as confirmed by the extremely similar distribution of spectral types at the top of the blue sequence. This age coincides with that found by \citet{beauchamp1994}, who used individual dereddening, but is younger than the values determined by both \citet{joshi2008} and \citet{subramanian2006}.

\subsection{Cluster mass}
\label{mass}

Our Str\"{o}mgren photometry covers the whole range of B-type stars, in which we found 179 cluster members. Using the individual dereddened absolute magnitudes, $M_{V}$, we assigned a mass to each cluster member, by calibrating against the isochrone of \citet{marigo2008}. Of course, this procedure is only valid 
for stars without emission lines.  Be stars cannot be individually dereddened and their apparent stellar parameters do not correspond to their actual mass \citep{fremat2005}. They tend to be brighter than non-emission stars of the same spectral type by a factor that depends on rotational velocity and inclination angle \citep[though they can be fainter when seen at high inclination; e.g.,][]{townsend2004}. Because the fraction of Be stars is very high in the cluster upper main-sequence, an accurate determination of the IMF is not possible, because the Be stars fall in bins with heavy weight.

We have tried to fit the IMF following the procedure of
\citet{maiz2005}, but our data cannot be adequately described by an
exponential function. There are three reasons contributing to
this. Firstly, our mass range is small and does not extend to
sun-like stars. Second, differential reddening is very likely
resulting in incompleteness for stars less massive than
$\sim5\,M_{\sun}$. Third, there is a real excess of stars in the
bins corresponding to $\sim8$ to $10\,M_{\sun}$. The same effect is
clearly seen in the IMF determination of \citet{beauchamp1994}, which shows a lack of 
stars with masses around $5\,M_{\sun}$ and an
excess around $8\,M_{\sun}$ (their Fig.~19). The lack of stars with
intermediate mass is also visible in Fig.~16 of \citet{joshi2008}. This
excess of stars around the turn-off might be caused by the large
number of Be stars, because they would appear brighter (and hence more
massive) than they really are because of two combined effects. On the
one hand, fast rotation will make them slightly brighter than slower
rotators of the same mass, as mentioned above. On the other hand, and
perhaps more importantly, the dereddening procedure will overestimate
their individual reddenings, because the circumstellar excess would be
taken for interstellar reddening. This over-correction of the
interstellar reddening should result in brighter absolute magnitudes. 

To correct for this possible effect, we repeated the IMF determination by
assigning the average cluster reddening to the Be stars and then
calculating $M_{V}$ with the $DM$ obtained from the isochrone fitting,
but found no improvement to the fit. We then repeated the
fit by adding 0.3~mag to their $M_{V}$ to correct for the brightening caused by fast rotation. 
In all cases, we still found a lack of stars around
$5\,M_{\sun}$ and an excess of stars around $8\,M_{\sun}$. Since our
photometry does not extend to lower masses and we cannot fully exclude that this effect is real, we did not try to determine the slope
of the IMF, and simply accepted the results from previous authors, who found  that it
approximates the Salpeter value when the full range to solar-like
stars is considered.

Our data, however, can be used to determine the total mass in B-type
stars contained in the cluster (including Be stars), which is
$M\approx1\,200\pm100\:M_{\sun}$, where the error bar reflects the
uncertainty in assigning a mass to the Be stars. This value includes
the B-type giants with an initial mass ($M_{{\rm ini}}$) close to $\approx14\,M_{\sun}$ (but not
the RSGs, which may have up to $M_{{\rm ini}}\approx14.5\,M_{\sun}$) and the
faintest objects in our sample with $\approx3\:M_{\sun}$. As
mentioned, we are probably not complete for stars less massive than $5\:M_{\sun}$. If we have $\approx 1\,000\:M_{\sun}$ between 5 and $14\:M_{\sun}$, where we are likely to be complete with 141 stars, a \citet{ms79} mass distribution implies $\sim 8\,000\:M_{\sun}$ for the whole cluster. A Kroupa-type IMF implies a slightly higher mass. These are only rough estimates, but, together with the high number of B-type stars, they suggest that NGC~7419 is a moderately massive cluster with a total mass in the $5-10\times10^{3}\:M_{\sun}$ range.  

\section{Discussion}
\label{dis}

We have presented for the first time Str\"omgren photometry and classification spectroscopy for B-type stars in NGC~7419. By using a large number of photometric standards in several young open clusters, we achieved a successful transformation to the standard system, even though the reddening to NGC~7419 is slightly higher than the reddening to any of our standard clusters. Since a correct transformation to the standard system requires the use of standard stars with similar intrinsic colours and similar reddenings to the target stars \citep{manfroid1987,crawford1994}, there was an {\it a priori} risk of finding systematic effects in the transformed colours. Fortunately, the transformation was fully successful, as demonstrated by the excellent match between photometric and spectroscopic spectral types (see Sect.~\ref{age}), showing that the standard clusters can be used to transform photometry of somewhat more reddened clusters.
 
Our distance estimate is higher than those of \citet{joshi2008} and \citet{subramanian2006}. This is almost entirely due to the younger age estimated for the cluster, because the difference in the absolute magnitude of the turn-off at 14~Myr and 24~Myr is $\approx$0.5 mag \citep{marigo2008}, resulting in a difference of $\approx$0.5 mag in $DM$. The observed spectral types fully support the younger age because no stars earlier than B2 should be observed at ages higher than 20 Myr. 

The radial velocity of NGC~7419 corresponds to $v_{{\rm LSR}}\approx$$-55\,{\rm km\,s^{-1}}$ \citep{beauchamp1994}. The dynamical distance for the cluster is thus $\sim4.5$~kpc for the Galactic rotation curve of \citet{reid2009}, which agrees quite well with our distance determination. There are, however, very significant deviations from the Galactic rotation curve in this region \citep{reid2009}. The open cluster NGC~7538 ($l=111\fdg54$, $b=0\fdg78$, $v_{{\rm LSR}}= -57\,{\rm km\,s^{-1}}$) has a parallax distance of $2.6\pm0.2$~kpc, but a kinematic distance of 4.6~kpc \citep{reid2009}. To reduce the distance to NGC~7419 to a value similar to the parallax distance to NGC~7538, we would have to assume a very substantial departure from the standard reddening law, which does not seem to be borne out by either our photometry or that from \citet{joshi2008}.    

\begin{figure}
  \centering
\resizebox{\columnwidth}{!}{\includegraphics[bb=22 32 691 516,clip]{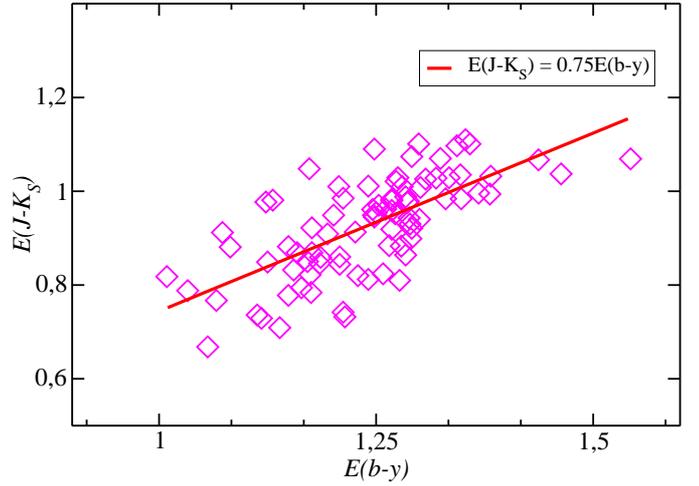}}
   \caption{$E(J-K_{{\rm S}})/E(b-y)$ diagram for normal B-type stars in the cluster. These points fit a straight line with a slope of 0.75. \label{Fig7}}
    \end{figure}

\citet{subramanian2006} assumed a standard reddening law with $R=3.1$, while \citet{joshi2008} calculated $R=3.2\pm0.1$. The dereddening procedure of \citet{crawford1970b} implicitly assumes a reddening law similar to that in the solar neighbourhood. To confirm its validity, we plot the individual $E(b-y)$ values for all B-type stars never classified as Be stars against their $E(J-K_{\rm S})$ excesses in Fig.~\ref{Fig7}. The infrared excesses were calculated using their 2MASS photometry and the intrinsic colours for the photometric spectral type from 
\citet{winkler1997}. Though the best-fit line, $E(J-K_{\rm S})=0.75\times E(b-y)$, is fully compatible within errors with a standard reddening law (for which values around 0.7 are expected for the slope of this relation), there is a large scatter around this line. Many stars have $E(J-K_{\rm S})\approx0.1$~mag higher or lower than expected for their $E(b-y)$. This scatter is most likely dominated by photometric uncertainties, because the uncertainties in $E(J-K_{\rm S})$ is $>0.04$, even for the best 2MASS magnitudes.  

\subsection{Be stars}

NGC~7419 presents the highest fraction of Be stars relative to the total number of B stars among all well-studied young open clusters in the Milky Way. The Be stars concentrate strongly above and around the main-sequence turn-off. \citet{subramanian2006} suggested that most of the Be stars in the cluster were Herbig Be stars instead of classical Be stars. This seems very unlikely if we consider the following: 
\begin{itemize}
\item Many Be stars present strong variability in their emission lines on time scales of years, including complete disappearance and reformation of the emission lines, a behaviour typical of classical Be stars and unreported in Herbig Be stars.
\item The observed equivalent width of H$\alpha$ is $\ga-50$\AA\ in all our spectra or those in \citet{malchenko2011}. These are values typical of early-type classical Be stars, while many Herbig Be stars present stronger emission.
\item As discussed by \citet{malchenko2011}, the Be stars in NGC~7419 present 
circumstellar excesses $E(J-K_{\rm S})$ in the range typical of classical Be stars in young open clusters. Most Herbig Be stars have much stronger excesses \citep{hillenbrand1992}.
\item The emission-line stars are very strongly concentrated towards the main-sequence turn-off, a characteristic typical of clusters containing classical Be stars \citep{mcSwain2005}. The distribution of Herbig Be stars, on the other hand, should follow the IMF, with the longer time-scale for disk dissipation favouring the detection of lower-mass systems. Moreover, a population of Herbig Be stars should be accompanied by the presence of lower mass pre-main sequence stars, unless the IMF is utterly anomalous. \citet{joshi2008} failed to detect any T Tauri star associated with the cluster.
\end{itemize} 

In summary, we find no evidence for Herbig Be stars in the cluster and accordingly assume that all emission stars are classical Be stars.

\subsection{Clusters with red supergiants}

\citet{clark2009b} carried out simulations of RSG populations by building synthetic clusters with masses of up to $4\times10^{4}\:M_{\sun}$ by populating them according to a \citet{kroupa01} IMF and evolving stars according to the solar metalicity evolutionary tracks of \citet{mm00} with high initial rotation ($v_{{\rm rot}}=300\:{\rm km}\,{\rm s}^{-1}$). According to these simulations, a cluster with 3\,000$\:M_{\sun}$ is expected to have $1\pm1$ RSG at an age 12-15~Myr, while a 10\,000$\:M_{\sun}$ cluster is expected to have $2\pm2$ RSGs. The probability that a 5\,000$\:M_{\sun}$ cluster has zero RSGs at 14~Myr is $>30$\%. 

These numbers seem compatible with observations if one considers that even moderately massive clusters such as NGC~869 \citep[$M_{\rm cl}\approx5\,000\:M_{\sun}$ in the core;][]{currie2010} or NGC~663 do not contain any RSGs in their core regions, while NGC~884 (slightly less massive than NGC~869) contains one in the core region and four in the halo\footnote{Though the corresponding mass of the halo is difficult to evaluate. One could alternatively state that there are ten RSGs in the whole of the Perseus Double Cluster complex, which contains at least $20\,000\:M_{\sun}$ \citep{currie2010}.}. The five RSGs in NGC~7419, with a mass $\la 10,000\:M_{\sun}$, lie outside the 1-$\sigma$ range derived from Monte-Carlo simulations, but well within the 2-$\sigma$ range, even though no other cluster with a well-determined mass and a similar age contains more than two RSGs \citep{eggenberger2002}.

We can thus assume that NGC~7419 represents the extreme tail of the statistical distribution.  After all, the simulations use a particular set of evolutionary tracks with somewhat extreme conditions (a very high initial $v_{{\rm rot}}$). However, the fact that it is also the Galactic cluster with the highest known fraction of Be stars in the upper main-sequence suggests the possibility of a physical reason for these phenomena.

The ratios of RSGs with respect to B-type stars and Be stars with respect to B stars seem to be correlated, and vary with metallicity \citep{meynet07}. The reasons for this correlation are unclear, because the ratio of RSGs to blue supergiants evolves with metallicity against theoretical expectations \citep{meynet11}. The metallicity of NGC~7419 is unlikely to be different from that of other clusters in its neighbourhood, though this should be checked with spectral modelling\footnote{Unfortunately, better (i.e., higher-resolution and higher-S/N) spectra than those displayed here are needed for this task.}. A high binary fraction may also have an impact on the number of RSGs, because stars in close binaries will undergo mass transfer during their evolution and never will grow to the size of a RSG.

Therefore we can propose a scenario where a cluster is formed with a low fraction of close binaries and thus angular momentum is not soaked up by binary motion, which results in very high initial $v_{{\rm rot}}$ for most stars. This scenario would explain the high fraction of Be stars and the high number of RSGs, because there would only be few binaries. However appealing such a scenario may be to explain the observed properties of NGC~7419, it cannot be taken as a rule for clusters with a high fraction of Be stars, as shown by the counterexample of NGC~663. This cluster, which has a total mass comparable to that of NGC~7419 \citep{pandey2005}, also has a very high Be fraction around the main-sequence turn-off \citep{negueruela05}. Its nuclear region, on the other hand, does not contain any RSG, but five blue supergiants of spectral types B and early A. Again, we might be probing the tail of the statistical distribution, but it is quite striking to note that two well-studied\footnote{NGC~7419 and NGC~663 are also sufficiently massive to be sure that the high fraction of Be stars is not due to small-number statistics. In both cases, we are considering $>150$ B-type stars.} clusters with similarly high fractions of Be stars present this widely different distribution of supergiants. In this sense, it is worthwhile noting that high rotational velocity can also be attained through mass transfer in a close binary \citep{marco2007,mcSwain2005},and so different physical mechanisms leading to the formation of Be stars may exist (see discussion in \citealt{mcSwain2005} or \citealt{tarasov2012}). 

It has also been claimed that the peculiarity of NGC~7419 is highly enhanced by the lack of any blue supergiant in the cluster, which makes its high number of RSGs still more unexpected \citep{beauchamp1994,caron2003}. The brightest blue star in the cluster, 103, was classified as B2.5\,II-III by \citet{caron2003}, based on a far-red spectrum. Our spectral type, based on a classification spectrum, is quite similar, BC2\,II, and places the object very close to the lower limit in luminosity for blue supergiants. A second object, 295, which had not been observed by previous authors because it is quite far away from the cluster core and very close to a much brighter foreground object, has a very similar spectrum, which we classify as BC2.5\,Ib-II, though noting that the slightly different classifications may be due to the different resolutions used. In the dereddened CMD (Fig.~\ref{Fig5}), these two objects have an $M_{V}$ slightly brighter than $-5$~mag, well above the cluster giants and $\sim$2~mag brighter than the MS turn-off. Another member with as similar absolute magnitude is the blue straggler star 180, BN\,0.5III. We can thus state that the lack of blue supergiants in NGC~7419 is partly a question of semantics, even though the ratio of RSGs to evolved blue stars is still extremely high when compared to other Galactic clusters \citep{eggenberger2002}, confirming its uniqueness.

\subsubsection{2MASS diagram}

The obscured clusters of red supergiants must be observed in the infrared, because heavy extinction renders them invisible at optical wavelengths. The line-of-sight to the clusters near the base of the Scutum arm is very complex, with many intervening populations that generate very heavy foreground contamination \citep{negueruela12}. The line-of-sight to NGC~7419 presents little contamination and the actual members can be selected from the Str\"omgren photometry. It is thus possible to create CMDs free from contamination to evaluate the effects of differential reddening and intrinsic dispersion.   

\begin{figure}
   \centering
\resizebox{\columnwidth}{!}{\includegraphics[clip]{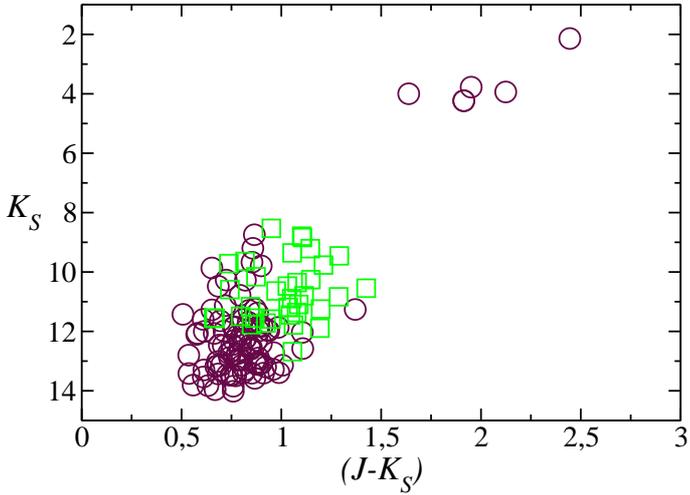}}
   \caption{$K_{{\rm S}}$ against $(J-K_{{\rm S}})$ for members of NGC~7419. Open squares represent Be stars in the cluster. 
\label{kjk}}
 \end{figure}

Figure~\ref{kjk} displays the 2MASS $K_{s}$/$(J-K_{{\rm S}})$ diagram for stars selected as members of NGC~7419 from the analysis of the Str\"{o}mgren photometry. The red supergiants are located on the upper right side of the diagram, clearly separated from the rest of the members that belong to the main sequence. The main sequence is very broad and dispersed, with $(J-K_{{\rm S}})$ ranging from $\approx$0.5 to 1.5. This dispersion is, to a large degree, caused by a significant population of Be stars that present high $(J-K_{{\rm S}})$ excesses because of circumstellar disks. The excesses observed amongst the Be stars in NGC~7419 range from $\sim$0 to $\sim$0.5, i.e., the same values as in field Be stars \citep{dougherty1994}. However, the non-emission-line stars also present a very wide range of values. Figure~\ref{Fig7} shows that the reddening follows a law close to the standard and thus the high dispersion must be caused by differential reddening, with $E(J-K_{{\rm S}})$ ranging from $\sim0.7$ to $\sim1.1$. 

The main sequence of NGC~7419 is difficult to trace in the $K_{{\rm S}}$/$(J-K_{{\rm S}})$ diagram even though we removed all foreground contamination and the reddening to the cluster is only moderate. For the obscured clusters of red supergiants, the reddening is much higher, the foreground contamination is much heavier, and the longer distance results in smaller spatial extent and hence frequent blending at the spatial resolution of 2MASS. It is thus easy to understand why the main-sequence populations of these clusters have not been identified from 2MASS data.

\subsubsection{Theoretical HR diagram for red supergiants}
	
\setcounter{table}{7}

We can use the cluster parameters found in our analysis to evaluate the accuracy of cluster parameters derived exclusively from the RSG population, as has been done for several clusters \citep[e.g.,][]{davies2007,clark2009a,negueruela10}.
In Table~\ref{IRSGs}, we present the 2MASS photometry for the five RSGs in NGC~7419. The errors in the determination of some magnitudes are high because of saturation problems, but we found no other broad-band near-infrared photometry for these stars in the literature. We derived the excesses in $(J-K_{{\rm S}})$ for each star taking the average intrinsic colours for each spectral type from \citet{carlos2012}. Assuming standard reddening, $A_{K}=0.67E(J-K_{{\rm S}})$, we calculated $K_{0}$ for each object. Then, we calculated $M_{K}$ for each RSG, using the distance modulus for the cluster, $DM=13.0$, determined in Section~\ref{distance} from the analysis of Str\"{o}mgren photometry. In obscured clusters, the distance is the most uncertain parameter, and must be estimated from the radial velocity  \citep[e.g.,][]{davies2007} or the run of extinction \citep[e.g.,][]{negueruela11}.

\begin{table*}
\caption{Photometric data for red supergiants in NGC~7419, from 2MASS.\label{IRSGs}}
\centering
\begin{tabular}{lcccccccc}
\hline\hline
Star&$J$&$E_{J}$&$H$&$E_{H}$&$K_{{\rm S}}$&$E_{K_{{\rm S}}}$&$E(J-K_{{\rm S}})$&$M_{K}$\\
\hline\hline
B921=56	&6.148&0.024&4.799&0.033&4.234&0.016&0.884&$-$9.358\\	
B696=122&5.726&0.018&4.662&0.212&3.775&0.036&0.891&$-$9.822\\		  
B139	&6.061&0.020&4.514&0.036&3.937&0.036&1.074&$-$9.783\\			
B950	&4.583&0.280&2.980&0.242&2.138&0.278&1.125&$-$11.616\\			
B435	&5.635&0.020&4.586&0.228&3.997&0.284&0.578&$-$9.390\\
\hline	
\end{tabular}				
\end{table*}

Finally, we used the following expressions, taken from \citet{levesque2005}, to calculate the intrinsic stellar parameters for the theoretical H-R diagram: 

\begin{equation}
(J-K)_{0}=3.10-0.547(T_{{\rm eff}}/1000\:{\rm K})
\end{equation}

\begin{equation}
 {\rm BC}_{K}=5.574-0.7589(T_{{\rm eff}}/1000\:{\rm K})
\end{equation}

\begin{equation}
(L/L_{\sun})=10^{-(M_{{\rm bol}}-4.74)/2.5}\, .
\end{equation}

The parameters determined are also listed in Table~\ref{IRSGs}.

\begin{figure}
\resizebox{\columnwidth}{!}{\includegraphics[clip]{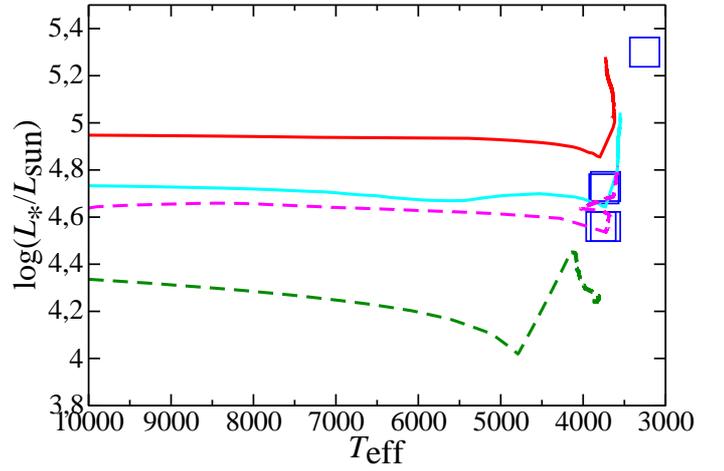}}
   \caption{HR diagram showing the locations of 5 RSGs in the cluster, with their positions derived from the spectral classification, assuming a distance to the cluster $d=4.0$~kpc. We also plot isochrones from \citet{meynet2000}. The solid lines are the $\log t=7.10$ (13 Myr; top, red), and $\log t=7.20$ (16 Myr; bottom, blue) isochrones with high initial rotation. The dotted lines are the $\log t=7.10$ (13 Myr; magenta), and $\log t=7.20$ (16 Myr; green) isochrones without rotation. The size of the symbols represents the errors in  $\log L_{*}$ that are due to observational uncertainties and calibration problems. The errors in $T_{{\rm eff}}$, though difficult to quantify, should be smaller than the symbol size.\label{Figteorica}}
    \end{figure}

In Figure~\ref{Figteorica}, we display the theoretical Hertzsprung-Russell (HR) diagram showing the positions of the five RSGs compared to different isochrones from \citet{meynet2000}. Their position appears to be consistent with the high rotation 16-Myr isochrone or the 13-Myr isochrone without rotation, but not with the 16-Myr isochrone without rotation. This is in excellent agreement with the 14-Myr age derived from the main-sequence turn-off, when we consider that a range of initial rotations is to be expected. In this theoretical HR diagram, the stars lie exactly at the expected positions for RSGs, confirming the idea that their somewhat awkward position in the $V/(b-y)$ diagram stems from the reddening procedure used for the isochrone\footnote{The reddening procedure uses a Cardelli law, a polynomial approximation that introduces artificial bumpiness at high reddening \citep{maiz2012}.}. The position of B950 lies beyond the end of the isochrone, at a much higher luminosity than that of the other four RSGs. This is not surprising given its exceptional nature (see next subsection), but we must also note that its 2MASS photometry is affected by very heavy saturation. In addition, this star presents strong mid-infrared excess \citep{fc74}, and it is possible that its $E(J-K_{{\rm S}})$ excess is slightly overestimated, resulting in a too bright luminosity ($L_{*}$). 

The excellent agreement of the theoretical isochrones with the positions of the RSGs provides further confirmation of the age determination. It also validates the use of similar diagrams to estimate the parameters of obscured clusters, for which the uncertainty in the distance clearly dominates all other sources of error. 
	  
\subsection{MY Cep as a standard for late-M supergiants}
\citet{fc74} remarked on the peculiarity of B950 (IRC +60 375; generally known by its variable star designation, MY~Cep), because supergiants later than M4 were not known in open clusters \citep{abt63}. Even such a luminous star as VY~CMa, which is at or past the Hayashi limit \citep{witt12}, has the spectral type M4--5\,Ia$^{+}$ (with some weak spectral variability). Confirmed supergiants with later spectral types are extremely rare \citep{wing09}. VX~Sgr shows pronounced variations in spectral type, from M4\,Ia to $\sim$M10\,Ia. \citet{solf} remarked that its spectrum was similar to those of late-M giants, but showing stronger metallic lines, and suggested that this might be the defining characteristic of late-M supergiants. The spectrum of MY~Cep seems to confirm this suggestion.

Another star with marked excursions in spectral type is S~Per, a member of the Per~OB1 association. This object spends most of its time as an M4\,Ia supergiant, but it has been observed with spectral type as late as M8 \citep[and references therein]{wing09}. This object offers a good comparison to MY~Cep, because the age of h \& $\chi$~Per is believed to be very close to that of NGC~7419, and these two stars should have similar masses. MY~Cep seems very unusual in that it permanently displays a late-M spectral type. Even though short excursions to other spectral types cannot be excluded, it seems to have varied by less than 0.5 spectral subtypes over the past $\sim60$~years. The only known star displaying a similar behaviour is NML~Cyg, which has been observed with a spectral type $\sim$M6\,Ia for the past $\sim50$~years \citep{wing09}. In view of this stability, NML~Cyg and MY~Cep could be taken as pseudo-standards for the classification of late-M supergiants, alleviating the lack of criteria discussed, among others, by \citet{negueruela12}.

All late-M supergiants are surrounded by extended envelopes. VY~CMa displays prominent ejecta that provides evidence for multiple and asymmetric mass loss events, and is surrounded by a dust envelope \citep{humphreys07}. An extended dust envelope has also been imaged around NML~Cyg \citep{schuster2009}. Maser mapping has allowed the detection of extended structures around S~Per and VX~Sgr and accurate distance determination for these sources via parallaxes \citep{asaki,chen}. Maser emission from MY~Cep has also been detected. Several surveys have found SiO, OH, and water masers \citep[and references therein]{verheyen}, indicative of heavy mass loss and an extended envelope. Its spectral energy distribution in the mid-IR is very similar to that of S~Per, though it suggests a smaller envelope \citep{fc74}. According to the Padova evolutionary tracks \citep{marigo2008}, MY~Cep, which is close to the end of its life at 14~Myr, should have a mass $\sim$14.5$\:M_{\sun}$ if it was born with a low rotational velocity.

\section{Conclusions}

By combining Str\"omgren photometry and classification spectra of its brightest stars, we have derived astrophysical parameters for the young open cluster NGC~7419. We obtained a distance of $\sim4$~kpc, compatible with the kinematic distance implied from its radial velocity and the rotation curve of \citet{reid2009}, and an age of $14\pm2$~Myr. The turn-off from the main sequence is found at spectral type B1, in excellent agreement with the age determination. The stellar population of NGC~7419 is similar to that of the double Perseus cluster, which has a similar age. We identified 179 B-type stars, covering the whole B1\,--\,B9 range, though differential reddening renders our sample incomplete for spectral types B6 and later. 

We calibrated the mass of our stars from their absolute magnitudes by mapping them onto the corresponding isochrone. We found 141 stars between 5 and 14$\,M_{\sun}$, implying a mass of $\sim 1\,000\,M_{\sun}$. Since previous authors have determined a mass distribution not far from Salpeter's, this implies an initial mass approaching $5\,000\,M_{\sun}$ in stars more massive than the Sun. Extrapolation to lower masses results in a total initial mass close to $10\,000\,M_{\sun}$, indicating that NGC~7419 is sufficiently massive to serve as a valid comparison for theoretical models and a template for the study of more obscured clusters. 

We found a large population of Be stars, strongly concentrated around the main-sequence turn-off. Indeed, above the turn-off there are more Be stars than non-emission stars. The fraction of Be stars is $\approx40\%$ above $V=17$ (types B2\,V and earlier) and very low for later types, confirming the trend seen in other clusters \citep{mcSwain2005}. Many of the Be stars in NGC~7419 have shown pronounced variability during the $\sim$10~years of monitoring, with episodes of disk loss and reformation being reported here for the first time. 

The 2MASS $K_{{\rm S}}$/$(J-K_{{\rm S}})$ diagram for NGC~7419 shows no clearly defined sequence, even if only confirmed members are plotted. This is due to the combined effects of differential reddening, with $E(J-K_{{\rm S}})$ for non-emission B-type stars ranging from $\approx0.7$ to $\approx1.1$~mag, and the circumstellar excesses of Be stars, which may add up to $\approx0.5$~mag to $E(J-K_{{\rm S}})$. The reddening law is close to standard, and the dereddened 2MASS magnitudes of the M-type supergiants may be used to calculate their intrinsic parameters, which provide a very good fit to theoretical tracks, demonstrating the validity of this procedure to estimate the properties of obscured clusters.

The presence of five red supergiants in a cluster with $\la 10\,000\,M_{\sun}$ is only marginally consistent with population synthesis models \citep{clark2009b}. When considered together with the very high fraction of Be stars and the very low ratio of evolved blue stars to RSGs, it strongly hints at the presence of some physical mechanism, perhaps related to high initial rotational velocities, as the source of these effects. Finally, MY~Cep has the latest spectral type of all known cluster RSGs, at M7.5\,I, and presents clear evidence for heavy mass loss. Unlike other extreme RSGs, it does not seem to have undergone any substantial spectral variability over the past 60 years, which makes it an important pseudo-standard to classify very late RSGs.

\begin{acknowledgements}

We thank the referee, Radostin Kurtev, for his helpful comments.

This research is partially supported by the Spanish Ministerio de
Ciencia e Innovaci\'on (MICINN) under
grants AYA2010-21697-C05-05 and CSD2006-70, and by the Generalitat Valenciana 
(ACOMP/2012/134).

 The WHT is operated on the island of La
Palma by the Isaac Newton Group in the Spanish Observatorio del Roque
de Los Muchachos of the Instituto de Astrof\'{\i}sica de Canarias. Partly based on observations made with the Nordic Optical Telescope, operated
on the island of La Palma jointly by Denmark, Finland, Iceland,
Norway, and Sweden, in the Spanish Observatorio del Roque de los
Muchachos of the Instituto de Astrof\'{\i}sica de Canarias.  Some of the data have been taken using ALFOSC, which is owned by the Instituto de Astrof\'{\i}sica de Andalucia (IAA) and operated at the Nordic Optical Telescope under agreement between IAA and the NBIfAFG of the Astronomical Observatory of Copenhagen. Based in part on observations made at Observatoire de Haute Provence (CNRS), France.

This research has made use of the Simbad database, operated at CDS,
Strasbourg (France) and of the WEBDA database, operated at the
Institute for Astronomy of the University of Vienna. This publication
makes use of data products from the Two Micron All Sky Survey, which is a joint project of the University of
Massachusetts and the Infrared Processing and Analysis
Center/California Institute of Technology, funded by the National
Aeronautics and Space Administration and the National Science
Foundation.
\end{acknowledgements}

\longtab{3}{
\begin{longtable}{lllllllllll}
\caption{Photometry and coordinates for stars in the field of 
NGC~7419.\label{tb4}}\\
\hline
Name&Name(WEBDA)&$\alpha$(J2000)&$\delta$(J2000)&$V$&$\sigma_{V}$&$(b-y)$&$\sigma_{(b-y)}$&$c_{1}$&$\sigma_{c_{1}}$&$N$\\
\hline\hline
\endhead
2  & 456&22 54 38.42&+60 45 33.80 &18.140&0.038        &1.089  &       0.049   &       0.994   &       0.088   &       1\\
4  &1595&22 54 10.55&+60 45 37.20 &17.515&0.037        &1.099  &       0.048   &       0.382   &       0.075   &       1\\
5  &1593&22 54 04.67&+60 45 40.31 &17.406&0.037        &1.167  &       0.048   &       0.537   &       0.076   &       1\\
6  &1594&22 54 22.27&+60 45 40.70 &16.209&0.007        &1.221  &       0.004   &       0.307   &       0.097   &       4\\
8  &1592&22 54 04.19&+60 45 43.00 &16.103&0.016        &0.672  &       0.019   &       0.224   &       0.056   &       4\\
9  & 390&22 54 31.55&+60 45 44.70 &18.187&0.037        &1.229  &       0.048   &       0.853   &       0.084   &       1\\
\hline
\end{longtable}
}

\longtab{4}{
\begin{longtable}{ccccccc}
\caption{ 2MASS photometry for member stars of 
NGC~7419.\label{irtab}}\\
\hline
Name&$J$&$\sigma_{J}$&$H$&$\sigma_{H}$&$K_{{\rm S}}$&$\sigma_{K_{{\rm S}}}$\\
\hline\hline
\endhead
\hline
&&&B-type stars&&&\\
\hline
2&	14.318	&	0.030	&	13.757	&	0.035	&	13.407	&0.038\\
4&	13.743	&	0.028	&	13.214	&	0.031	&	12.964	&0.033\\
5&	12.851	&	0.036	&	12.250	&	0.039	&	11.867	&0.035\\
6&	12.113	&	0.021	&	11.529	&	0.022	&	11.244	&0.020\\
9&	14.003	&	0.031	&	13.380	&	0.031	&	13.119	&0.034\\
\hline
&&&Supergiants&&&\\
\hline	
56&	6.148	&	0.024	&	4.799	&	0.033	&	4.234	&0.016\\
122&	5.726	&	0.018	&	4.662	&	0.212	&	3.775	&0.036\\
B139&	6.061	&	0.020	&	4.514	&	0.036	&	3.937	&0.036\\
B435&	5.635	&	0.020	&	4.586	&	0.228	&	3.997	&0.284\\
B397&	4.583	&	0.280	&	2.980	&	0.242	&	2.138	&0.278\\
\hline
&&&Be type stars&&&\\
\hline	
18&	11.503	&	0.021	&	10.887	&	0.020	&	10.473	&0.019\\
30&	13.086	&	0.023	&	12.373	&	0.020	&	11.893	&0.020\\
34&	12.642	&	0.020	&	12.052	&	0.022	&	11.719	&0.019\\
44&	10.454	&	0.020	&	9.958	&	0.020	&	9.717	&0.019\\
47&	12.442	&	0.019	&	11.747	&	0.022	&	11.245	&0.020\\
\hline
\end{longtable}
}

\longtab{7}{
\begin{longtable}{cccc}
\caption{Photometry for membership of NGC~7419.\label{photparam}}\\
\hline
Name&$E(b-y)$&$M_{V}$&Spectral Type(photometric)\\
\hline\hline
\endhead
2&1.131&0.278&(B8--B9)\,V\\				  
4&	1.201	&	$-$0.648	&	B1.5\,V			\\			  
5&	1.255	&	$-$0.990	&	(B2.5--B3) \,V		\\				  
6&	1.334	&	$-$2.526	&	B1\,V			\\			  
9&	1.287	&	$-$0.349	&	(B7--B8) \,V		\\				  
\hline
\end{longtable}
}
\end{document}